\documentclass[usenatbib]{mn2e}
\input{psfig.sty}
\usepackage{graphicx}
\usepackage{dcolumn}
\usepackage{float}
\usepackage{threeparttable}

\title[Cyclotron Resonance Energies and Orbital Elements of 4U 0115+63]
{Cyclotron Resonance Energies and Orbital Elements of Accretion Pulsar 4U 0115+63 During the Giant Outburst in 2008}

\author[Jun Li, Wei Wang and Yongheng Zhao]{Jun Li$^{1},^{2},^{3}$\thanks{E-mail:JunLi.leon@gmail.com},
Wei Wang$^{1}$\thanks{E-mail:wangwei@bao.ac.cn} and Yongheng Zhao$^{1},^{3}$ \\
$^{1}$National Astronomical Observatories, Chinese Academy of Sciences, Beijing 100012, China\\
$^{2}$Graduate University of Chinese Academy of Sciences, Beijing 100012, China\\
$^{3}$Key Laboratory of Optical Astronomy, National Astronomical Observatories, Chinese Academy of Sciences, Beijing 100012, China\\}
\begin{document}

\date{}

\pagerange{\pageref{firstpage}--\pageref{lastpage}} \pubyear{2011}

\maketitle

\label{firstpage}

\begin{abstract}
In this paper, we present both timing and spectral analysis of the
outburst of 4U 0115+63 in April -- May 2008 with INTEGRAL and RXTE observations.
We have determined the spin period of the neutron star at $\sim 3.61430 \pm 0.00003$ s, and a spin up
rate during the outburst of $\dot{P}=(-7.24 \pm 0.03)\times10^{-6} {\rm s~d^{-1}}$, the angle of periapsis
 $\omega=48.67^\circ \pm 0.04^\circ$ in 2008 and its variation (apsidal motion) $\dot{\omega} = 0.048^\circ \pm 0.003^\circ {\rm yr}^{-1}$
when comparing with previous measured values of $\omega$. We also confirm the relation of spin-up torque versus luminosity in this source during the giant outburst.
The hard X-ray spectral properties of 4U 0115+63 during the outburst are studied with INTEGRAL and RXTE. Four cyclotron absorption lines (fundamental at $\sim 10-15$ keV, 2nd harmonic at $\sim 22$ keV,
3rd at $\sim 34$ keV and 4th at $\sim 45$ keV) are detected
using the spectra from combined data of IBIS and JEM-X aboard INTEGRAL in the energy range of 3 -- 100 keV. Two cyclotron absorption lines are measured from the spectra combining PCA and HEXTE data in the energy band of 4 -- 50 keV.
The 5 -- 50 keV luminosities at an assumed distance of 7 kpc are determined to be
in the range of $(1.5-12)\times 10^{37} {\rm ergs~ s^{-1}}$ during the outburst. The fundamental absorption line
energy varies during the outburst: around 15 keV during the rising phase, and transiting to $\sim 10$ keV during
the peak of the outburst, and further coming back to $\sim 15$ keV during the decreasing phase. The variations of
 photon index show the correlation with the fundamental line energy changes: the source becomes harder around the
 peak of the outburst and softer in both rising and decreasing phases. This correlation and transition processes
during the outburst need further studies in both observations and theoretical work. The known relation of the
fundamental line energy and X-ray luminosity is confirmed by our results, however, our discoveries suggest that some other factors besides luminosity play the important role in fundamental line energy variations and spectral transitions.

\end{abstract}


\section{Introduction}

The X-ray transient 4U 0115+63 is one of the best studied X-ray sources with different instruments. It is a high mass X-ray binary composed of a
neutron star and a Be star. Ever since it was first observed in the $Uhuru$ satellite
survey (Giacconni et al. 1972, Forman et al. 1978), series of observations have been done to
study its different kinds of features. With the precise positional determinations by the satellites
 \textsl{SAS-3, Ariel V and HEAO-1}, the companion star was identified as a heavily reddened O9e star, V635 Cas, with visual
magnitude $V \approx 15.5$ (see Cominsky et al. 1978, Johnston et al. 1987, Johns et al. 1978,
Hutchings \& Crampton 1981). The Be star is an early type, non-supergiant star which some times exhibits
Balmer series lines. The distance of the star is suggested to be $\sim 7$ kpc (Negueruela \& Okazaki 2001). The spin period of neutron star in 4U 0115+63 was discovered at $\sim 3.6$ s (Cominsky et al. 1978).
The binary's orbital elements were first determined by Cominsky et al. (1978), Rappaport et al. (1978), and Kelly et al. (1981),
using the $SAS-3$ data sets, finding the eccentric($e=0.34$), and orbital period 24.3 days, and precise timing analysis
in different time intervals have given a upper limit on the apsidal motion $\dot{\omega} \leq 2.1^\circ {\rm yr}^{-1}$.

This hard X-ray emission is produced via accretion process by a neutron star accreting material from
its companion Be star. However, due to different physical and geometrical conditions in the accretion
disc, the source has shown different types of X-ray activities. In quiescence, 4U 0115+63 exhibits low
luminosity($L_{x} \leq 10^{34} {\rm ergs~s}^{-1}$) (Campana et al. 2001). Some times the object presents Type I X-ray
outbursts ($10^{36} {\rm ergs~s}^{-1}\leq L_{x}\leq 10^{37} {\rm ergs~s}^{-1}$) separated by orbital period.
For this source, the type II outbursts appear more often then the Type 1.
It shows giant X-ray outburst($L_{x} \geq 10^{37} {\rm ergs~s}^{-1}$) enduring for
several weeks or months. The onset of the Type II outbursts occur near periastron, and some
times it is understood as quasi-periodic behaviors with a modulation period of about 3 to 5 years.
The physical origin of the outburst mechanism is still unclear. Negueruela \& Okazaki (2001) suggest that
 the distance at which the circumstellar disc is truncated depends on the orbital elements and viscosity.
 Binary system with low eccentricity would expect the gap between the disc outer radius and the lobe radius
of the Be star so large that normally the neutron star could not accrete enough gas at periastron passage
to express Type I outburst. In this case, these systems would only display occasional Type II outburst.

Research about cyclotron resonance spectral features (CRSFs) of this source has been carried out
repeatedly with different observations since CRSF was clearly detected as an absorption feature at $\sim 20$ keV (Wheaton 1979) using the data from HEAO-1 A4. Soon after that, White et al. (1983) pointed out that the CRSF at $\sim 20$ keV in fact was the second harmonic resonance, and two cyclotron absorption lines at the fundamental energy $\sim$ 11.5 keV and 23 keV were detected. The double harmonic absorption
features were confirmed using \textsl{Ginga} data (Nagase et al. 1991; Tamura et al. 1992). Heindl et al. (1999) discovered a third absorption line at $\sim 33.56$ keV in 1999 with RXTE. Moreover, a fourth absorption line was detected by Santangelo et al. (1999) at 49.5 keV with \textsl{BeppoSAX} and Tsygankov et al. (2007) at 44.93 keV with INTEGRAL. Recently the fifth harmonic resonance absorption line at $\sim 53$ keV was reported by RTXE observations (Ferrigno et al. 2009). So that, 4U 0115+63 is the only object in the spectrum of which five cyclotron line harmonics
were detected, making it as one of the best objects for studying the physics of
cyclotron resonance in the polar cap regions of X-ray pulsars in binaries.

4U 0115+63 was observed at different X-ray luminosity levels, and the fundamental energy of CRSFs was repeatedly measured around 11 keV around the peak of the Type II outbursts. But the variations of the fundamental line energy were detected when we observed the source in the lower X-ray luminosity. The report from Ginga (Mihara 1995; Mihara et al. 1998) found that during the outburst in 1991, when the X-ray luminosity is about 7 times lower than the typical luminosity of the outburst peaks, instead
of the familiar double absorption features at $\sim 11$ and 22 keV, the
outburst exhibited a single deep and wide absorption at $\sim 16$ keV. The further studies suggested that the 11 keV fundamental energy line moves to $\sim 16$ keV in lower luminosity ranges instead of that the 22 keV second harmonic decreasing to 16 keV (Nakajima et al. 2006). Detailed studies
showed a luminosity-dependent change of the fundamental absorption line energy (Burnard et al. 1991; Nakajima et al. 2006; Tsygankov et al. 2007).

4U 0115+63 undergone the giant Type II outbursts frequently about every 3--5 years. The latest giant outbursts occurred in  in April -- May 2008. During this giant outburst, both INTEGRAL and RXTE have carried out pointing observations on the source frequently. In this paper, we analyzed both INTEGRAL and RXTE data of 4U 0115+63 obtained during giant outbursts and studied the temporal and spectral properties and variations during the outbursts. We will concentrate on the spin properties of the neutron star, deriving the orbital elements of the binary and the resonance absorption energy variations during the outburst.

\section[]{Observations}

\begin{table}
\caption{Observations of 4U 0115+63 during the outburst in 2008 with INTEGRAL and RXTE,
the instrument, observed date, exposure time are shown.}

\begin{center}
\scriptsize
\begin{tabular}{c c c l}
\hline \hline
Instrument & MJD (Rev.) & Exposure time (s)\\
\hline
JEM-X & 54546.52(664) & 99599.08\\
JEM-X & 54555.46(667) & 49290.03\\
JEM-X & 54558.45(668) & 39747.02\\
JEM-X & 54563.38(669) & 50285.74\\
JEM-X & 54566.41(670) & 46669.91\\
JEM-X & 54573.90(673) & 52292.43\\
JEM-X & 54580.62(675) & 110950.9\\
\hline
IBIS  & 54546.52(664) & 99599.08\\
IBIS  & 54555.46(667) & 50067.22\\
IBIS  & 54558.45(668) & 51651.26\\
IBIS  & 54563.38(669) & 53380.74\\
IBIS  & 54566.41(670) & 49267.91\\
IBIS  & 54573.90(673) & 52293.74\\
IBIS  & 54580.62(675) & 110952.9\\
\hline
\hline
Instrument & MJD (pointing) & Exposure time (s)\\
\hline
HEXTE &	54545.265(93032010100) &	13984\\
HEXTE &	54545.470(93032010101) &	1680\\
HEXTE &	54548.074(93032010200) &	16928\\
HEXTE &	54549.186(93032010201) &	11424\\
HEXTE &	54549.971(93032010202) &	14864\\
HEXTE &	54550.236(93032010203) &	4256\\
HEXTE &	54546.970(93032010204) &	20640\\
HEXTE &	54551.083(93032010205) &	7584\\
HEXTE &	54551.284(93032010206) &	2336\\
HEXTE &	54556.254(93032010300) &	9936\\
HEXTE &	54556.142(93032010301) &	3088\\
HEXTE &	54554.226(93032010302) &	13584\\
HEXTE &	54558.932(93032010303) &	15392\\
HEXTE &	54560.043(93032010400) &	16096\\
HEXTE &	54562.011(93032010401) &	14816\\
HEXTE &	54562.945(93032010402) &	12944\\
HEXTE &	54563.986(93032010403) &	8384\\
HEXTE &	54564.817(93032010404) &	15504\\
HEXTE &	54565.864(93032010405) &	21344\\
HEXTE &	54566.975(93032010406) &	2208\\
HEXTE &	54567.001(93032010500) &	1376\\
HEXTE &	54567.956(93032010501) &	16160\\
HEXTE &	54569.983(93032160100) &	15520\\
HEXTE &	54571.944(93032160101) &	15616\\
HEXTE &	54574.037(93032160200) &	9680\\
HEXTE &	54578.360(93032160201) &	13136\\
\hline
PCA &	54545.265(93032010100) &	13984	\\
PCA &	54545.470(93032010101) &	1680	\\
PCA &	54548.074(93032010200) &	16928	\\
PCA &	54549.186(93032010201) &	11424	\\
PCA &	54549.971(93032010202) &	14864	\\
PCA &	54550.236(93032010203) &	4256	\\
PCA &	54546.970(93032010204) &	20640	\\
PCA &	54551.083(93032010205) &	7584	\\
PCA &	54551.284(93032010206) &	2336	\\
PCA &	54556.254(93032010300) &	9936	\\
PCA &	54556.142(93032010301) &	3088	\\
PCA &	54554.226(93032010302) &	13584	\\
PCA &	54558.932(93032010303) &	15392	\\
PCA &	54560.043(93032010400) &	16096	\\
PCA &	54562.011(93032010401) &	14816	\\
PCA &	54562.945(93032010402) &	12944	\\
PCA &	54563.986(93032010403) &	8384	\\
PCA &	54564.817(93032010404) &	15504	\\
PCA &	54565.864(93032010405) &	21344	\\
PCA &	54566.975(93032010406) &	2208	\\
PCA &	54567.001(93032010500) &	1376	\\
PCA &	54567.956(93032010501) &	16160	\\
PCA &	54569.983(93032160100) &	15520	\\
PCA &	54571.944(93032160101) &	15616	\\
PCA &	54574.037(93032160200) &	9680	\\
PCA &	54578.360(93032160201) &	13136	\\
\hline
\end{tabular}
\end{center}

\end{table}

The giant outburst of 4U 0115+63 occurring in 2008 was observed by INTEGRAL(\textsl{INTEernational Gamma-Ray Astrophysics
Laboratory}, Winkler et al. 2003) and RXTE(\textsl{Rossi X-ray Timing Explorer}, Bradt et al. 1993).

INTEGRAL is ESA's currently operational space-based hard X-ray/soft gamma-ray telescope.
We mainly use the data collected with the softer gamma-ray imager (IBIS/ISGRI; Lebrun et al. 2003)
 and the small X-ray telescope JEM-X (Lund et al. 2003). IBIS has a 12'(FWHM) angular resolution and arcmin
source location accuracy in range of energy at 18 -- 200 keV, JEM-X obtains a lower energy band
 at 3 -- 35 keV. Two instruments
are co-aligned, allowing simultaneous observations in a wide
energy range.

RXTE is a space X-ray telescope run by
NASA. The pointing observations are performed mainly by two instruments: Proportional Counter Array (PCA) and High Energy X-ray Timing Experiment (HEXTE). PCA is a spectrometer with an effective area of 6400 $cm^{2}$, with energy
range of 2--60 keV and time resolution of 1 microsec. HEXTE covers a higher energy range of 15 -- 200 keV with a timing resolution of 8 microsec. The two instruments aboard RXTE are very effective for timing analysis on fast variability of X-ray sources.

In Table 1, we summarize the archival data of 4U 0115+63 during the outburst in 2008 with both INTEGRAL and RXTE observations, and the instrument, date and exposure
time are listed in details.

The used INTEGRAL data are available from
the INTEGRAL Science Data Center (ISDC). The analysis was done with the standard INTEGRAL off-line
scientific analysis (OSA, Goldwurn et al. 2003) software, ver. 9.0. INTEGRAL observed the outburst in seven revolutions (one INTEGRAL orbit, 3 days). Each revolution includes several individual pointing observations (one science window) lasting from about 3000 s to 10000 s.
Individual pointings in each satellite revolution processed with OSA 9.0 were mosaicked to create sky images for the source detection. For each revolution, we got one spectrum from 3 -- 100 keV using the combined spectra of JEM-X and IBIS. IBIS has a poor time resolution which is larger than 1 second, not suitable for timing analysis of the 3.6 sec X-ray pulsar in 4U 0115+63. In timing analysis, only JEM-X data are used to search for spin period of the neutron star in 4U 0115+63. The light curves of JEM-X have a time resolution of 0.5 s.

Data of RXTE are available from
the standard products of the RXTE data are available on RXTE archive, which are used for the light curve extractions with a time resolution of 0.125 s from both of PCA and HEXTE. Each science window records data with the durations of several thousand seconds (see Table 1). The spectrum for each science window is extracted using the combined data of PCA and HEXTE.

\section{DATA ANALYSIS AND RESULTS}

\subsection{TIMING ANALYSIS AND ORBITAL ELEMENTS OF 4U 0115+63}

The spin properties of the accretion pulsars are crucial for us to understand
the accretion physics and giant outburst mechanisms. The determination of spin period and period derivatives of the X-ray pulsars in binaries will sensitively depend on the orbital parameters. In addition pulse profile variations of X-ray pulsars during the outbursts is not well understood and requires more analysis. In this subsection, we will study the temporal properties of 4U 0115+63 during the outburst of 2008 using the INTEGRAL and RXTE data. We will try to derive the spin period of the X-ray pulsar combined with the orbital parameters. Study the apsidal motion of X-ray binaries is one of the important features to understand the nature
of the X-ray binaries' properties, so we also try to search for the apsidal motion of 4U 0115+63 by comparing the results in different time intervals.

X-ray light curves of 4U 0115+63 during the 2008 outburst by JEM-X are extracted in two energy bands: 3-- 10 keV and 10 -- 35 keV, with a time-resolution of 0.5 s. The light curves of PCA are used in the energy band of $2 -40$ keV.

A general way to perform timing analysis in determining the pulse period of X-ray pulsars in binaries is to find the pulse arrival time (see Rappaport et al. 1978, Tamura et al. 1992).
The pulse arrival time for each data set was determined by comparing the folded
pulse profile with a reference profile. This method can get the exact arrival time of each pulse of
each data set. However, since the pulse profile of X-ray pulsar would change during the outburst, the determined arrival time of the position of pulse profile (e.g., peak) may have large uncertainties, it needs additional statistical test to ensure the results. In this paper,
we have carries out another way to do the timing analysis on pulse period studies (see Raichur \& Paul 2010). It would not be affected by the variation of the pulse profile.

This method is mainly based on the HEAsoft, published by NASA which aims
to handle the high energy data set obtained from the space satellite, such as INTEGRAL, RXTE.
First of all, we corrected the times of each photon to the barycenter of the solar system, then
 we used EFSEARCH (a build-in function in HEAsoft) to calculate the estimated spin
period of the source in each data set with best value based on Chi-Square estimation.
The EFSEARCH might depend on the epoch we used, so we could correct it as follows:
calculate the $\chi^2$ with different epochs, $T(epoch) = t_0 + 0.1\,i\,P/N$, where $i$ = 0, 1, 2, ..., 9,
and N is the number of phase bins in pulse profile. Then we average ten sets of $\chi^2$ versus period, fit the peak
with quadratic and find the maximum which should be the estimated period. The 1$\sigma$ error could be given
when the quadratic is lowered by 1.0.

As we get all the pulse periods of data sets during the outburst, we fit the data according to the doppler motion
of this source. The observed pulse period is given as below:

\begin{equation} \label{eq:Pspin}
P_{spin}^{obs}(t) = (P_{spin}(t_0) + \dot{P}_{spin}(t - t_0)) \sqrt{\frac{1 + \upsilon_r/c}{1 - \upsilon_r/c}}.
\end{equation}
Here $\dot{P}_{spin}$ denotes the derivative of spin period, $P_{spin}^{obs}$ is the
spin period values observed at time $t\ {\rm and}\ t_0$ is the reference time of the outburst. $\upsilon_r$
is the component of velocity along the sight of neutron star which depends on the orbital
elements:
\begin{equation} \label{eq:upsilon}
\upsilon_r = \frac{2 \pi a_x\sin i}{(1-e^2)^{1/2} P_{orb}}(\cos(\nu + \omega) + e \cos \omega),
\end{equation}
the $a_x \sin i$ denotes for the projected semi-major axis, P$_{orb}$ is the orbital period,
$\omega$ is the angle of periastron and $e$ is the eccentricity of the neutron star's
orbit. $\nu$ is the true anomaly and is closely related to the periastron passage time
$T_{\omega}$, the relationship could be given as follows:
\begin{equation} \label{eq:nu}
\tan \frac{\nu}{2} = \tan \frac{E}{2} \sqrt{\frac{1 + e}{1 - e}},
\end{equation}
\begin{equation} \label{eq:E}
E - e\sin E = \frac{2 \pi}{P_{orb}}(t - T_{\omega}).
\end{equation}
Here $E$ stands for the eccentric anomaly. Combining all those equations, we could get
the doppler motion of the neutron star in the different orbital phases by fitting the observed spin period values in different time intervals.

Using the EFSEARCH tool, we derive the observed values of spin period by INTEGRAL/JEM-X and RXTE/PCA from MJD 54545 -- 54580
in the 2008 outburst. The observed data points of the spin period during the outburst are presented in Fig. 1.
These data points are fitted by the function described by Eq. (1). The fitting is based on Levenberg-Marquardt algorithm
for nonlinear least square method, weighted by error of each point. The parameters' errors are calculated from the covariance matrix
of the parameter estimates. We adopted some orbital parameters from Bildsten et al. (1997), i.e., the orbital period
$P_{orb} = 24.317037 \pm 0.000062$ day, $a_x\sin i = 140.13 \pm 0.08$ lt-sec.
Then, we finally get the spin period, the derivative and other orbital elements of 4U 0115+63 as shown in Table 2.

The spin period of the neutron star is determined at $\sim 3.61430 \pm 0.00003\,s$ at the day MJD 54566,
with the derivative of $\dot P\sim (-7.24 \pm 0.03)\times10^{-6}$ s day$^{-1}$, and $T_{\omega} =$ MJD $54531.7709 \pm 0.0603 $.
The neutron star in 4U 0115+63 still undergone the spin-up process
during the 2008 giant outburst, which is consistent with the previous results on the history outbursts.
Here we summarize previous results on the spin period and period derivative of 4U 0115+63 in
Table 3 for a comparison. From 1981 -- 2008, the derived spin period values show the long-term spin-up trend. During the
history outbursts, 
during the outburst in 2008, $\dot{P}/P \sim -7.31 \times 10^{-4}$ yr$^{-1}$, which is considerably larger than before.

\begin{table}
\caption{Spin period and orbital elements of 4U 0115+63 at MJD 54566}
\centering
 \begin{threeparttable}
\begin{tabular}{l c}
\hline
\hline
Parameter & Value \\
\hline

$P$\,(s) (at MJD 54566) & 3.61430 $\pm$ 0.00003 \\
$\dot{P}_{spin}$ ($s\,day^{-1}$) & $(-7.24 \pm 0.03)\times10^{-6}$\\
$\omega\,(^\circ)$ & $48.67 \pm 0.04$\\
$\dot{\omega}\,(^\circ yr^{-1})$ & $0.048 \pm 0.003$ \\
$T_{\omega}\,(MJD)$ & 54531.7709\,$\pm$\,0.0603 \\
$P_{orb}\,(d)$\, \tnote{(1)} & 24.317037 $\pm$ 0.000062 \\
$e$\tnote{(1)} & 0.3402 $\pm$ 0.0002\\
$a_x\sin i\, (lt-sec)$\tnote{(1)} & 140.13 $\pm$ 0.08\\
Reduced $\chi^2 (d.o.f.)$ & 0.85 (68)\\

\hline
\end{tabular}
\begin{tablenotes}
 
  \scriptsize{\item[(1)]{By referring the results in Bildsten et al. (1997)}}
\end{tablenotes}
\end{threeparttable}
\end{table}


\begin{table*}
\caption{History of spin period and its derivative of 4U 0115+63 }
\centering
 \begin{threeparttable}
\begin{tabular}{c c c c }
\hline
\hline
MJD & $P_{spin}$ (s)& $\dot{P}/P$ (yr$^{-1})$ & References \\
\hline
40963 & 3.614658 $\pm$ 0.000036 & $(-3.3 \pm 1.4) \times 10^{-6}$ & Kelley et al. (1981) \\
42283 & $3.6142 \pm 0.0001$ & $-3.4 \times 10^{-5}$\, \tnote{(1)} &  Whitlock et al. (1989)\\
43540 & $3.6145737\pm 0.0000009$ & $(-3.2\pm 0.8)\times 10^{-5}$ & Rappaport et al. (1978)\\
44589 & 3.6146643 $\pm$ 0.0000018 & $(-2.6 \pm 0.5) \times 10^{-4}$ & Ricketts et al. (1981) \\
47941 & 3.614690 $\pm$ 0.000002 & $(-1.8 \pm 0.2) \times 10^{-4} $ & Tamura et al. (1992) \\
49481 & 3.6145107 $\pm$ 0.0000010 & $(-1.8 \pm 0.6)\times 10^{-4} $ & Scott et al. (1994) \\
50042 & 3.614499 $\pm$ 0.000004 & $(-4.2 \pm 0.9) \times 10^{-5} $ & Finger et al. (1995) \\
50307 & 3.614451 $\pm$ 0.000006 & - & Scott et al. (1996) \\
51232 & 3.614523 $\pm$ 0.000003 & - & Wilson et al. (1999) \\
51240 & 3.61447 $\pm$ 0.00002 & $(-2.52 \pm 0.6) \times 10^{-4}$ & Raichur \& Paul (2010) \\
53254 & 3.61436 $\pm$ 0.00002 & $(-3.84 \pm 0.55) \times 10^{-4}$ & Raichur \& Paul (2010) \\
54566 & 3.61430 $\pm$ 0.00003 & $(-7.31 \pm 0.03)\times10^{-4}$ & this work \\
\hline
\end{tabular}
\begin{tablenotes}
 
  \scriptsize{\item[(1)]{By directly comparing previous spin period in 1971 and 1974}}
  \end{tablenotes}
\end{threeparttable}
\end{table*}

\begin{table*}
\caption{Epoch and $\omega$ history of 4U 0115+63 }
\centering
 \begin{threeparttable}
\begin{tabular}{c c c c c}
\hline
\hline
Orbit number & Periastron time passage (TJD) & $\omega\ (^\circ)$ & Satellite & References \\
\hline
0 & 40963.08 $\pm$\,0.17 & $51.10\,\pm\,3.60$ & \textsl{Uhuru} & Kelly et al. (1981b)\\
106 & 43540.451 $\pm$\,0.006 & $47.66\,\pm\,0.17$ & \textsl{SAS-3} & Rappaport et al. (1978)\\
149 & 44585.700\, \tnote{(1)} & $47.15\,\pm\,0.$13 & \textsl{Ariel-6} & Ricketts et al. (1981)\\
287 & 47941.530 $\pm$\,0.006 & $48.02\,\pm\,0.$11 & \textsl{Ginga} & Tamura et al. (1992)\\
342 & 49279.2677$\pm$\,0.0034 & $47.66\,\pm\,0.$09 & \textsl{BATSE} & Cominsky et al. (1994)\\
422 & 51224.6465$\pm$\,0.051 & $48.50\,\pm\,0.$92 & \textsl{RXTE} & Raichur \& Paul (2010)\\
505 & 53243.038$\pm$\,0.051 & $50.07\,\pm\,1.86$ & \textsl{RXTE} & Raichur \& Paul (2010)\\
558 & 54531.7709$\pm$\,0.0603 & $48.67\,\pm\,0.04$ & \textsl{INTEGRAL\&RXTE} & this work\\
\hline
\end{tabular}
\begin{tablenotes}

  \scriptsize{\item[(1)]{The orbit number and $T_{\omega}$ is derived from Ricketts et al. (1981)}}
  \end{tablenotes}
\end{threeparttable}

\end{table*}

\begin{figure}
\centering
\includegraphics[angle=0,width=0.45\textwidth]{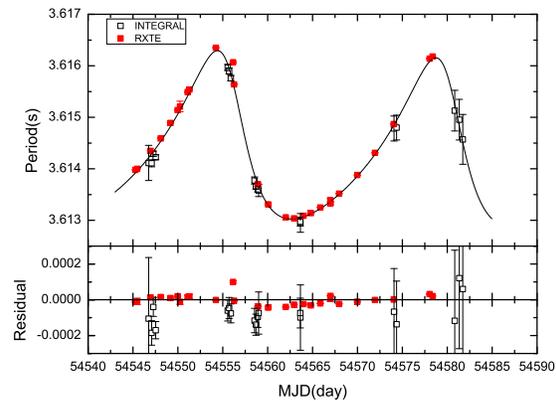}
\caption{Spin period variations due to doppler motion of neutron star in 4U 0115+63
during the 2008 outburst. The black squares denote for the spin period measured by INTEGRAL/JEM-X,
and the solid squares by RXTE/PCA. The residuals are shown in the lower
panel. The solid curve represents the expected doppler motion of the binary
due to the best fitted orbital elements.}
\end{figure}


We also try to measure the apsidal motion of the binary system. Using the INTEGRAL and RXTE
data in the 2008 outburst, we have derived the angle of periastron $\omega\sim 48.67^\circ \pm 0.04^\circ$ in
2008. We collected  the previous measurements of the angle of periastron of 4U 0115+63 in different epoches in Table 4. In Fig. 2,
the angle of periastron $\omega$ versus the observed date is plotted. The angle of periastron seems to change in the last 30 years,
suggesting existence of the apsidal motion. A linear function is used to fit the data points, also using Levenberg-Marquardt method
 and derive a apsidal motion rate
 $\dot{\omega} = 0.048^\circ \pm 0.003^\circ$ yr$^{-1}$, error is calculated from the covariance matrix of the linear fit,
  $R^2$ of this fit is 0.86.

\begin{figure}
\centering
\includegraphics[angle=0,width=0.55\textwidth]{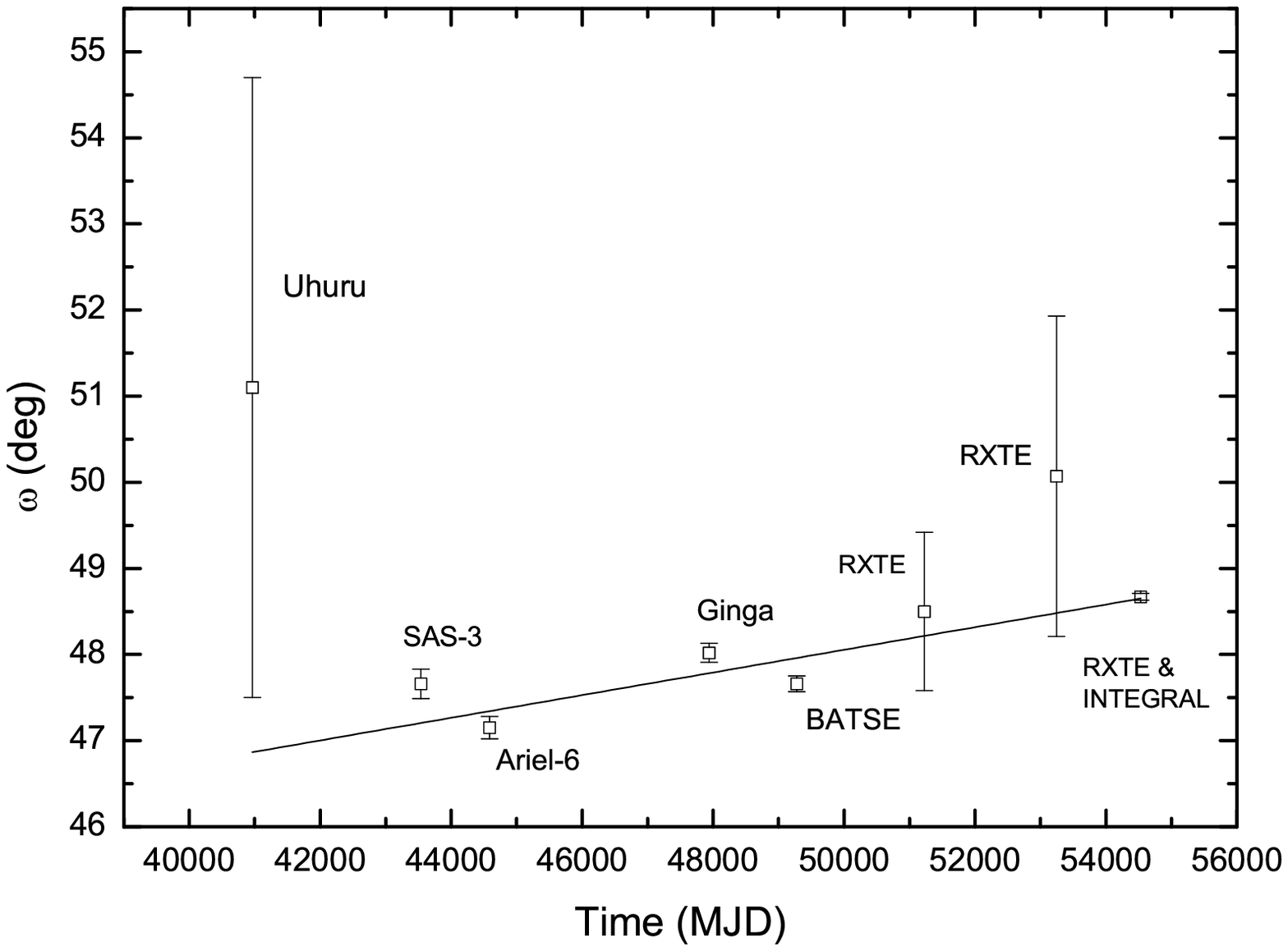}
\caption{The measured $\omega$ values of 4U 0115+63 from 1971 to 2008. A linear function is used to fit the data points in deriving average $\dot{\omega}\sim 0.048^\circ\pm 0.003^\circ$. }
\end{figure}

In addition, we also have studied the evolution of pulse profile of 4U 0115+63 during this giant outburst.
In Figure. 3, we displayed the pulse profiles with the outburst time in two energy bands of JEM-X: 3 -- 10 keV and 10 --35 keV. There exist obvious different features in pulse profiles of the high energy band and
low energy band. In the lower band, pulse profiles show secondary peaks in most time of the giant outburst, and may disappear near the end of the outburst (MJD 54581.7).
As the flux arrives at its peak, the secondary peak's amplitude also increases, so that we would see
the secondary peak clearly (see the upper panel of Figure 3, MJD 54558.6). As the outburst flux decreases, the secondary peak
also starts to merge with the main peak.
While at high energy band, the second peak is not so obvious. After the outburst reaches its
peak, the secondary peak could be seen (see lower panel of Figure 3, MJD  54574.0).
This difference may be due to the different mechanisms of pulsation at different
energy levels.

\begin{figure*}
\centering
\includegraphics[angle=0,width=0.45\textwidth]{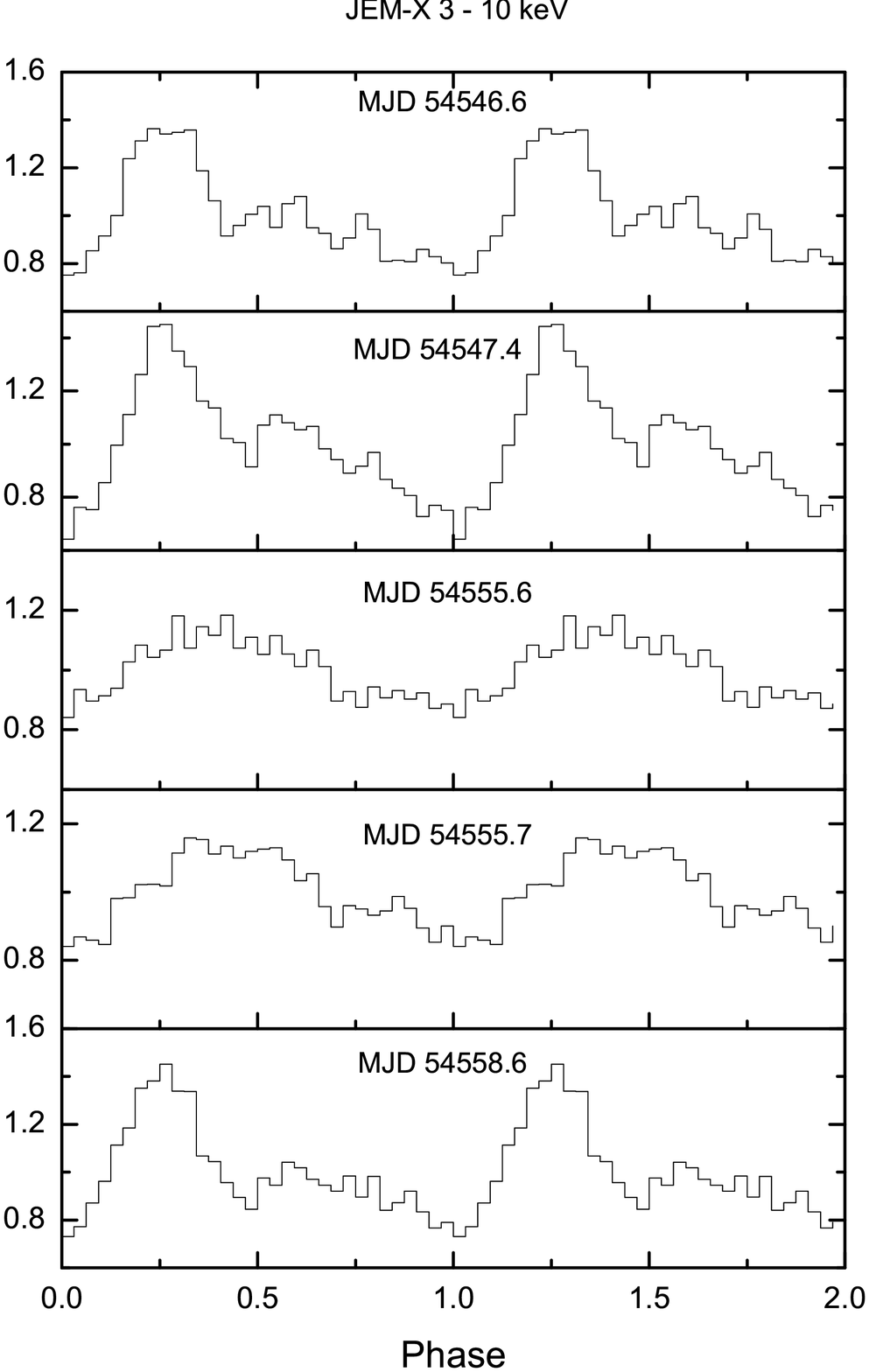}
\includegraphics[angle=0,width=0.45\textwidth]{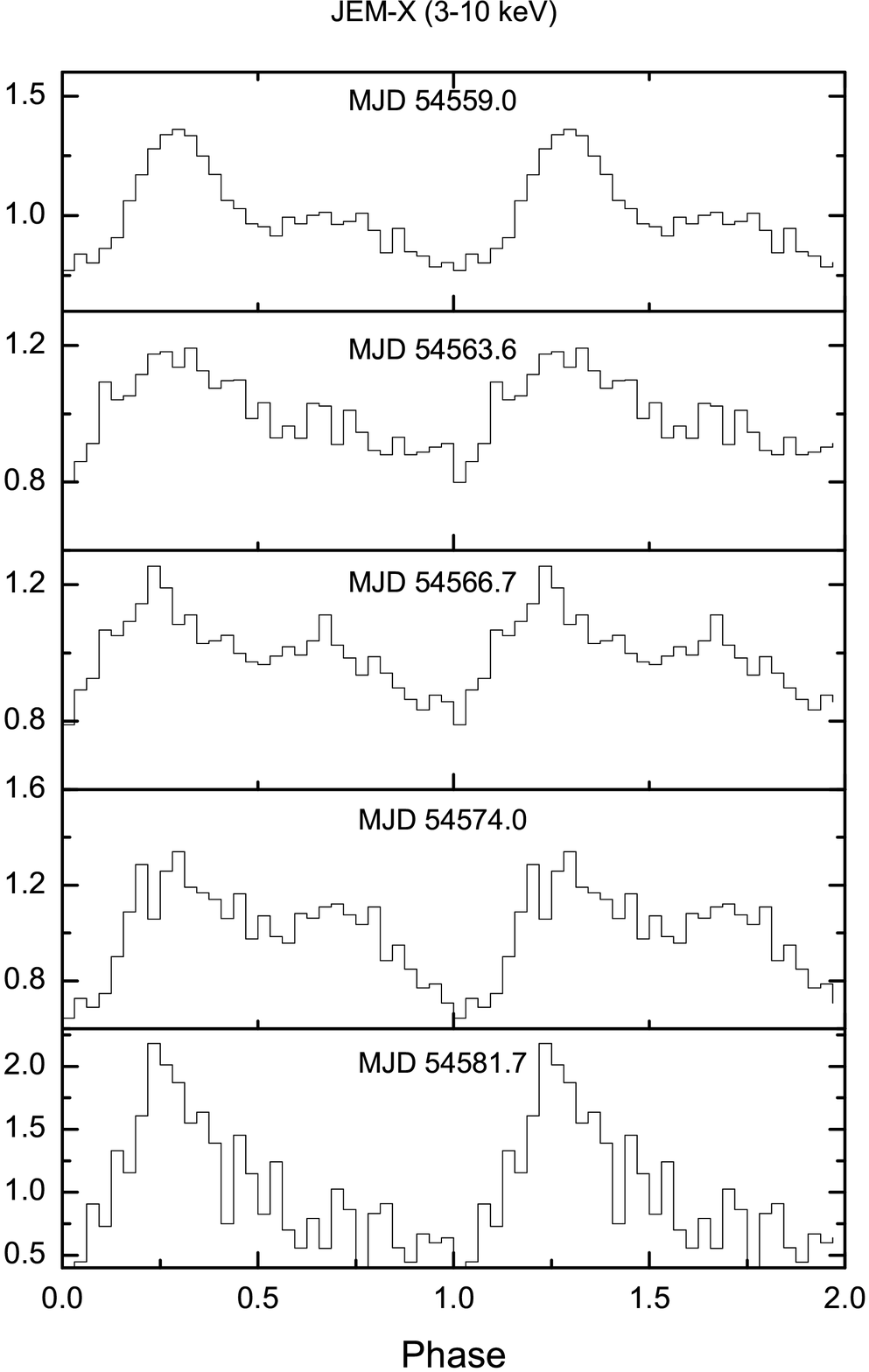}
\includegraphics[angle=0,width=0.45\textwidth]{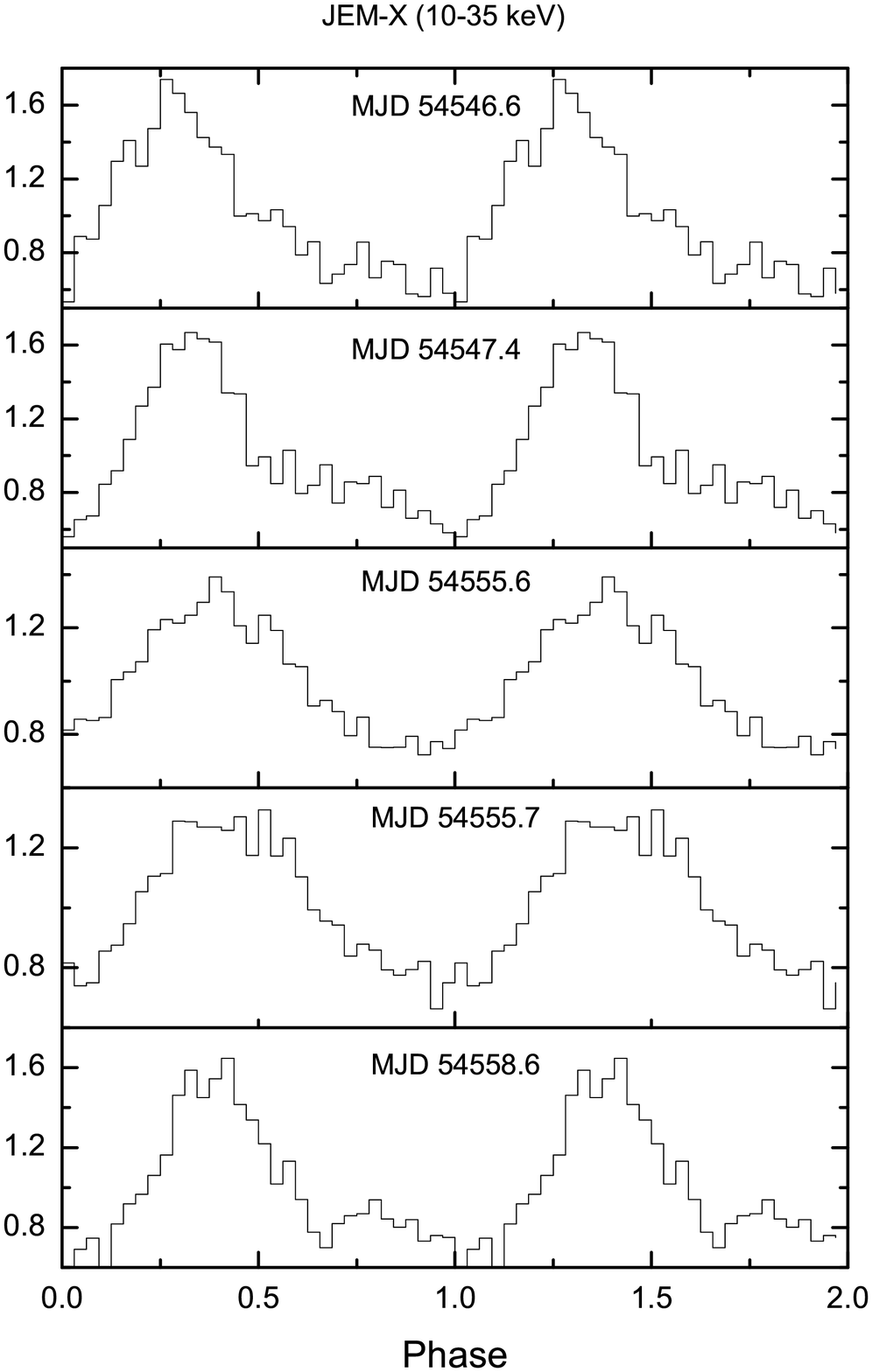}
\includegraphics[angle=0,width=0.45\textwidth]{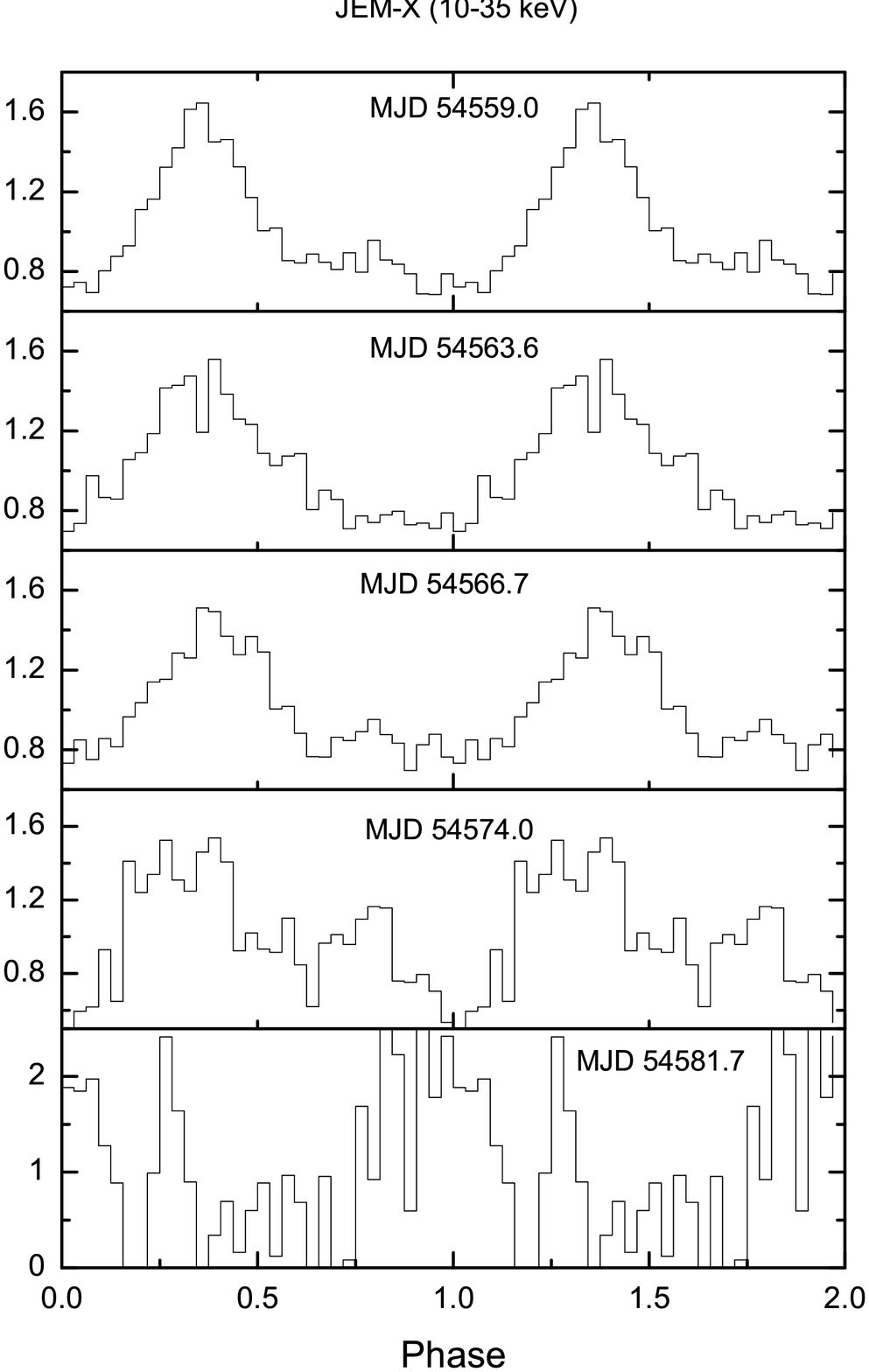}
\caption{The evolution of pulse profiles of 4U 0115+63 during the outbursts observed by INTEGRAL/JEM-X in two energy bands: 3--10 keV and 10--35 keV, respectively}
\end{figure*}

\begin{figure*}
\centering
\includegraphics[angle=0,width=0.45\textwidth]{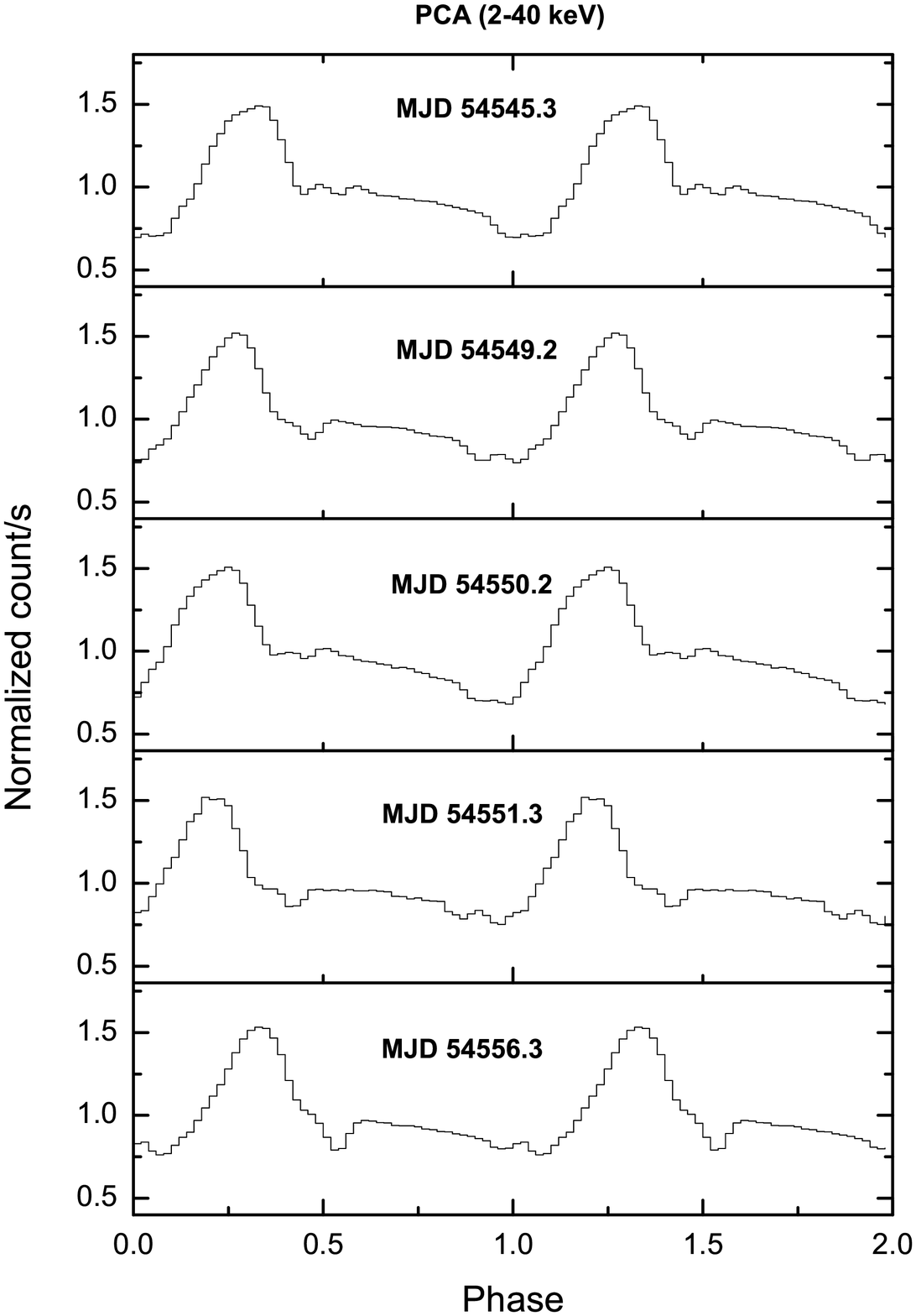}
\includegraphics[angle=0,width=0.45\textwidth]{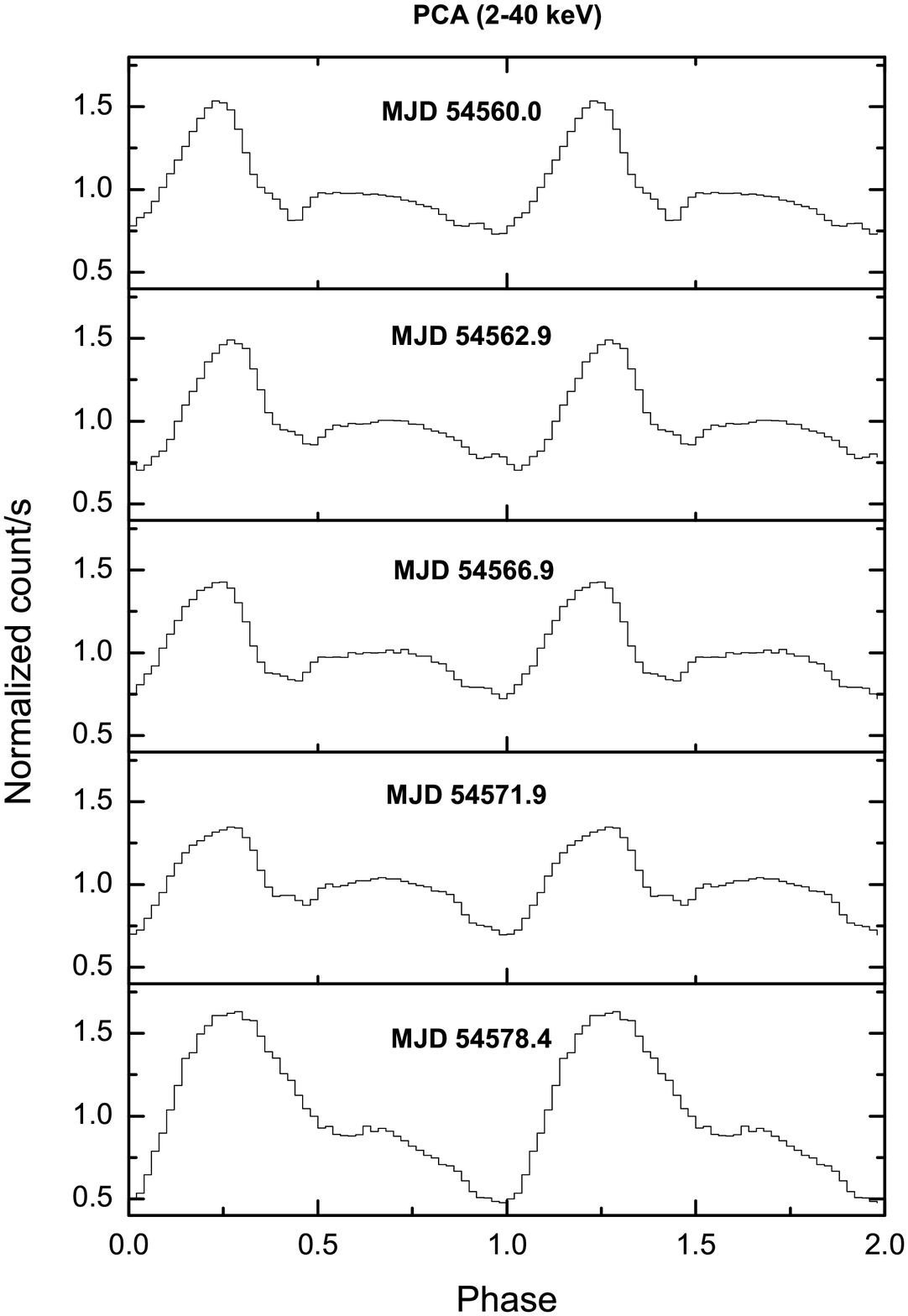}
\caption{The evolution of pulse profiles of 4U 0115+63 observed by RXTE/PCA during the outburst in the energy band of 2 -- 40 keV.}
\end{figure*}

In Fig. 4, we also presented the pulse profiles with the outburst evolution by RXTE/PCA in the energy band of 2 -- 40 keV. PCA has a better time resolution (0.125 s) and wider energy band, that could show us a vivid picture of the
evolution of pulse profile during the outburst. The second peak was detected in all time intervals, and it gradually
grows up, and becomes more obvious around the peak of the outburst,
then it begins to fade, which is consistent with the pulse profile evolution detected by INTEGRAL/JEM-X. The last pulse profile in MJD 54578.4 (Figure 4) almost merges with the main peak.

\subsection{Spectral Analysis and Cyclotron Absorption Features}

In this section we performed the spectral analysis of the hard X-ray spectra from several keV
up to 100 keV with RXTE and INTEGRAL respectively. We ignored spectral data with poor
signal to noise ratio in several energy channels, and used the spectral data as follows:(a) JEM-X: 3--31 keV; (b) IBIS: 28-100 keV; (c)
PCA: 4--30 keV; and (d) HEXTE: 17 -- 50 keV.

We fit the data using the XSPEC package. All the spectra have been fitted with a power-law type spectra
modified by a high energy cut off structure (the powerlaw*highecut model in
XSPEC package; White et al. 1983), additionally modified by cyclotron absorption lines in Lorenz
profile (see Mihara et al. 1990):
$\exp\left[ -D_f \frac{(W_f E/ E_{cycl})^2}{(E-E_{cycl})^2+W_f^2}\right]$
 with a fixed iron line (gaussian profile) at 6.4 keV (see Tsygankov et al. 2007).
Besides, photo-electric absorption (see Morrison and McCammon 1983)
 was added when fitting the INTEGRAL spectral data. Hard X-ray flux is measured in the range of 4--50 keV for
 both satellites.

\begin{table*} \label{table:integral spec}

\caption{Spectral Parameters of 4U 0115+63 with INTEGRAL}
\centering
\begin{threeparttable}
\begin{tabular}{c c c c c c c}
\hline
Model Parameters           & Rev 664 & Rev 667 & Rev 668 & Rev 669 & Rev 670 & Rev 673\\
\hline
$N_H (10^{22} cm^{-2}$) & - & 1.75\,$\pm$1.23 &  13.49\,$\pm$2.67 & - & -& - \\
Photo Index & 0.65\,$\pm$0.02 & 0.15\,$\pm$ 0.32 & 0.20\,$\pm$ 0.36  & 0.65\,$\pm$0.02 & 0.62\,$\pm$0.02 & 0.69\,$\pm0.03$\\
$E_{cut}\,(keV)$ & 9.17\,$\pm$0.14 & 9.13\,$\pm$0.30 &8.23\,$\pm$0.77 & 9.16\,$\pm$0.14 & 8.41\,$\pm$0.17 & 8.42\,$\pm$0.23 \\
$E_{fold}\,(keV)$ & 12.48\,$\pm$0.75 & 11.71\,$\pm$1,24 & 13.13\,$\pm$2.51 & 12.53\,$\pm$0.72 & 11.43\,$\pm$0.39 & 10.37\,$\pm$ 0.96 \\

$E_{cycl,1}\,(keV)$ & 16.86\,$\pm$0.63 & 9.53\,$\pm$0.62 & 10.03\,$\pm$1.27 & 16.54\,$\pm$0.60 & 14.65\,$\pm$0.47 & 15.62\,$\pm$0.73 \\
F-test Probability & 3.87$\times10^{-84}$ & 1.32$\times10^{-149}$& 5.343$\times10^{-54}$ & 6.249$\times10^{-70}$ & 7.566$\times10^{-83}$ & 1.336 $\times10^{-46}$ \\
$Depth_{cycl,1}$ & 0.57\,$\pm$0.14 & 0.66\,$\pm$0.15 & 0.73\,$\pm$0.31 & 0.52\,$\pm$0.12 & 0.33\,$\pm$0.11 & 0.58\,$\pm$0.18 \\
$Width_{cycl,1}\,(keV)$ & 5.76\,$\pm$1.18 & 5.25\,$\pm$0.68 & 5.23\,$\pm$1.45 & 5.41\,$\pm$1.15 & 4.44\,$\pm$1.20 & 5.11\,$\pm$1.52 \\

$E_{cycl,2}\,(keV)$ & 22.93\,$\pm$0.54 & 20.57\,$\pm$0.31 & 20.46\,$\pm$0.46 & 22.62\,$\pm$0.59 & 20.97\,$\pm$0.99 &  22.62\,$\pm$0.90  \\
F-test Probability & 0.0019 & 1.09$\times10^{-24}$& 2.377$\times10^{-39}$& 0.0017 & 0.0037 & 5.914 $\times10^{-8}$ \\
$Depth_{cycl,2}$ & 0.62\,$\pm$0.17 & 1.07\,$\pm$0.12 & 1.09\,$\pm$0.20 &1.12\,$\pm$0.04 & 0.74\,$\pm$0.01 & 1.10\,$\pm$0.21 \\
$Width_{cycl,2}\,(keV)$ & 3.42\,$\pm$1.02 & 7.51\,$\pm$0.59 & 6.90\,$\pm$0.82 & 3.87\,$\pm$0.88 & 6.14\,$\pm$0.85 & 4.10\,$\pm$1.25 \\

$E_{cycl,3}\,(keV)$ & 34.26\,$\pm$0.49 & 36.69\,$\pm$0.33 & 35.30\,$\pm$0.11 & 34.32\,$\pm$0.48 & 35.09\,$\pm$0.76 & 35.40\,$\pm$1.17 \\
F-test Probability &  0.0057 & 5.213$\times10^{-8}$ &  3.421$\times10^{-6}$&  0.0004 & 0.0008 & 0.0425 \\
$Depth_{cycl,3}$ & 0.57\,$\pm$0.14 & 0.58\,$\pm$0.09 & 0.38\,$\pm$0.11 & 0.56\,$\pm$0.14 & 0.50\,$\pm$0.16 & 0.98\,$\pm$ 0.38  \\
$Width_{cycl,3}\,(keV)$ & 1.00 (fixed) & 1.00(fixed) & 1.00(fixed) & 1.00(fixed) & 1.00(fixed) & 1.00(fixed) \\

$E_{cycl,4}\,(keV)$ & 43.97\,$\pm$2.07 & 46.50\,$\pm$2.16 & 43.74\,$\pm$5.58 & 43.96\,$\pm$1.64 & 43.56\,$\pm$2.07 & 46.05\,$\pm$1.58 \\
F-test Probability &  0.0001 & 7.521$\times10^{-20}$ & 9.667$\times10^{-22}$&  0.0001 & 0.0005 &  0.0297 \\
$Depth_{cycl,4}$ & 0.73\,$\pm$0.13 & 1.11\,$\pm$0.23 & 1.38\,$\pm$0.43 & 0.74\,$\pm$0.14 & 0.41\,$\pm$0.01 & 1.35\,$\pm$0.35  \\
$Width_{cycl,4}\,(keV)$ & 13.90\,$\pm$4.26 &20.48\,$\pm$4.38 & 24.87\,$\pm$7.50 & 14.03\,$\pm$4.33 &  8.49\,$\pm$4.54 & 9.96\,$\pm$5.79 \\


$E_{Fe}$\,(keV) (fixed) & 6.4 & 6.4 & 6.4 & 6.4 & 6.4 & 6.4\\
$Width_{Fe}\,(keV)$ (fixed)& 0.2  & 0.2 & 0.2 & 0.2 & 0.2 & 0.2\\
flux($10^{-9} ergs\,cm^{-2}s^{-1}$)\, \tnote{(1)} & 5.96 & 12.89 & 11.05 & 9.97 & 8.24 & 3.04 \\
$\chi^{2}\,(d.o.f)$ & 2.16(178) & 2.19(182) & 1.63(183) & 2.16(178) & 2.23(181) & 1.69(157)\\
\hline
\end{tabular}
\begin{tablenotes}
 \scriptsize
  \scriptsize{\item[(1)]{range: $5\sim50$ keV}}
  \end{tablenotes}
\end{threeparttable}

\end{table*}

\begin{table*} \label{table:rxte spec}

\caption{Spectral Parameters of 4U 0115+63 with RXTE}
\centering
\begin{threeparttable}

\begin{tabular}{c c c c c c c}
\hline
Model Parameters           & 93032010101 & 93032010206 & 93032010301 & 93032010406 & 93032010500 & 93032060201\\
\hline

Photo Index & 0.32 $\pm$ 0.01 & 0.40 $\pm$ 0.01 & 0.16 $\pm$ 0.09 & 0.29 $\pm$ 0.01 & 0.31 $\pm$ 0.01 & 0.56 $\pm$ 0.04 \\
$E_{cut}\,(keV)$ & 8.97 $\pm$ 0.05 & 8.42 $\pm$ 0.04 & 8.39 $\pm$ 0.08 & 8.20 $\pm$ 0.06 & 8.30 $\pm$ 0.07 & 9.09 $\pm$ 0.08 \\
$E_{fold}\,(keV)$ & 8.01 $\pm$ 0.18 & 8.47 $\pm$ 0.07 & 8.06 $\pm$ 0.23 & 8.42 $\pm$ 0.14 & 8.29 $\pm$ 0.19 & 8.93 $\pm$ 0.54 \\

$E_{cycl,1}\,(keV)$ & 15.42 $\pm$ 0.23 & 15.23 $\pm$ 0.11 & 10.04 $\pm$ 0.79 & 14.34 $\pm$ 0.20 & 14.66 $\pm$ 0.25 & 15.55 $\pm$ 0.20 \\
F-test Probability & 1.569$\times10^{-20}$ & 6.188$\times10^{-14}$ & 2.560$\times10^{-26}$ & 5.313$\times10^{-36}$ & 1.31$\times10^{-33}$ & 1.926$\times10^{-19}$ \\
$Depth_{cycl,1}$ & 0.30 $\pm$ 0.06 & 0.43 $\pm$ 0.02 & 0.41 $\pm$ 0.04 & 0.32 $\pm$ 0.05 & 0.28 $\pm$ 0.09 & 0.16 $\pm$ 0.09 \\
$Width_{cycl,1}\,(keV)$ & 2.84 $\pm$ 0.46 & 4.97 $\pm$ 0.32 & 7.25 $\pm$ 0.76 & 3.98 $\pm$ 0.45 & 3.67 $\pm$ 0.72 & 2.22 $\pm$ 1.13 \\

$E_{cycl,2}\,(keV)$ & 20.27 $\pm$ 0.63 & 22.31 $\pm$ 0.39 & 21.70 $\pm$ 0.21 & 20.20 $\pm$ 0.70 & 19.82 $\pm$ 1.20 & 19.98 $\pm$ 1.55 \\
F-test Probability & 0.0001 & 1.333$\times10^{-5}$ & 1.966$\times10^{-5}$ & 4.049$\times10^{-6}$ &  0.002 & 2.261$\times10^{-6}$ \\
$Depth_{cycl,2}$ & 0.64 $\pm$ 0.05 & 0.24 $\pm$ 0.03 & 0.39 $\pm$ 0.02 & 0.53 $\pm$ 0.04 & 0.52 $\pm$ 0.07 & 0.79 $\pm$ 0.09 \\
$Width_{cycl,2}\,(keV)$ & 5.29 $\pm$ 0.73 & 3.96 $\pm$ 0.69 & 4.59 $\pm$ 0.47 & 6.64 $\pm$ 0.75 & 6.62 $\pm$ 1.08 & 8.43 $\pm$ 1.87 \\

$E_{Fe}$\,(keV) (fixed) & 6.4 & 6.4 & 6.4 & 6.4 & 6.4 & 6.4\\
$Width_{Fe}\,(keV)$ (fixed)& 0.2  & 0.2 & 0.2 & 0.2 & 0.2 & 0.2\\

flux($10^{-9} ergs\,cm^{-2}s^{-1}$)\, \tnote{(1)} & 7.06 & 15.23 & 18.60 & 10.46 & 9.68 & 2.19 \\
$\chi^{2}\,(d.o.f)$ & 1.65 (57) & 1.98 (57) & 1.99 (57) & 1.58 (57) & 1.41 (71) & 1.06 (57) \\

\hline
\end{tabular}
\begin{tablenotes}
 \scriptsize
  \scriptsize{\item[(1)]{range: $5\sim50$ keV}}
  \end{tablenotes}
\end{threeparttable}
\end{table*}

\begin{figure*}
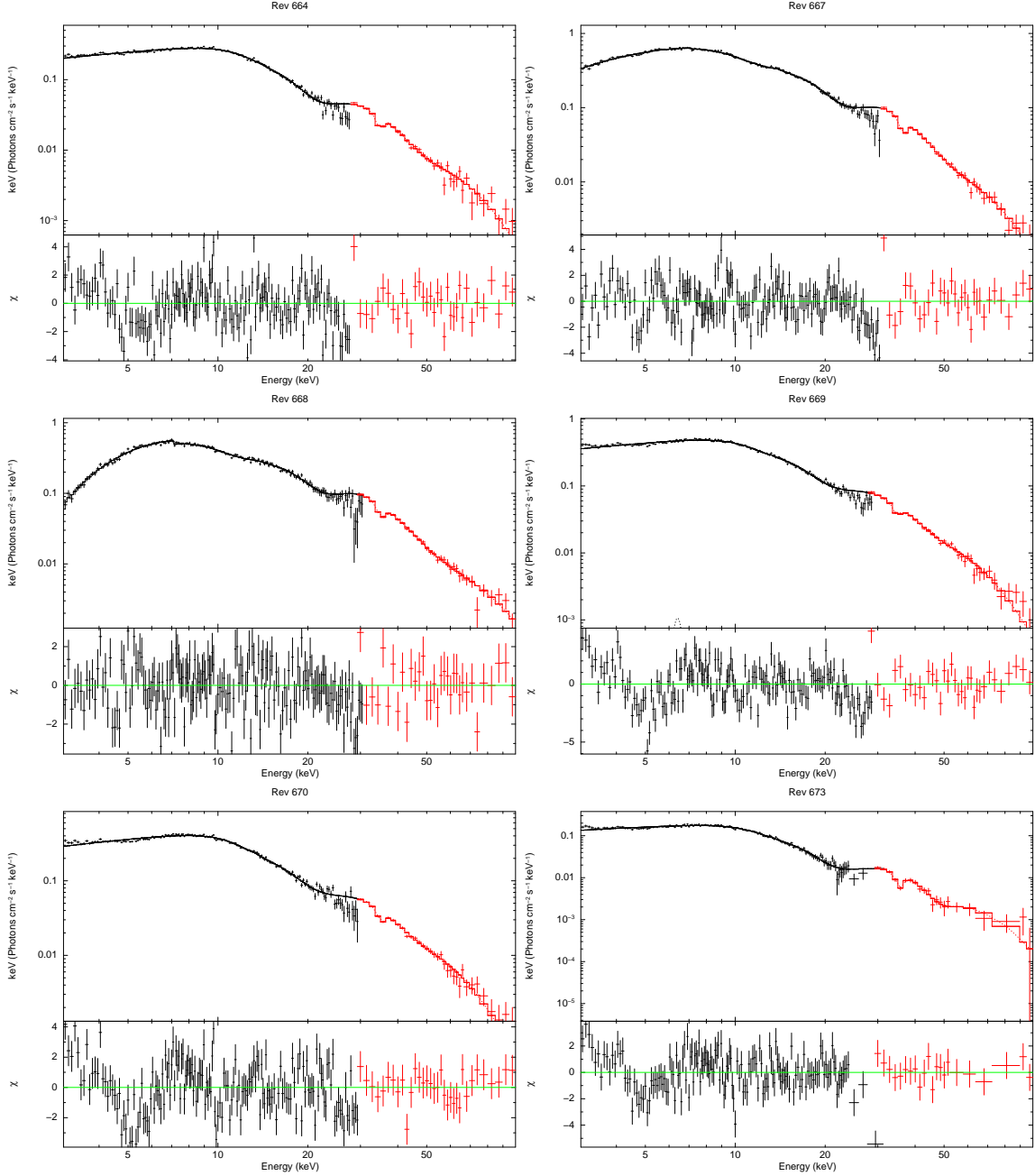

\centering
\includegraphics[angle=270,width=0.425\textwidth]{spectrum/jmx664_4th.ps}
\includegraphics[angle=270,width=0.425\textwidth]{spectrum/jmx667_4th.ps}
\includegraphics[angle=270,width=0.425\textwidth]{spectrum/jmx668_4th.ps}
\includegraphics[angle=270,width=0.425\textwidth]{spectrum/jmx669_4th.ps}
\includegraphics[angle=270,width=0.425\textwidth]{spectrum/jmx670_4th.ps}
\includegraphics[angle=270,width=0.425\textwidth]{spectrum/jmx673_4th.ps}
\caption{The hard X-ray spectra of 4U 0115+63 obtained by INTEGRAL/JEM-X (the red line) and
 INTEGRAL/IBIS (the black line) during the observation in revolution 664, 667,
 668, 669, 670, 673, the parameters are listed in Table 5}
\end{figure*}

\begin{figure*}
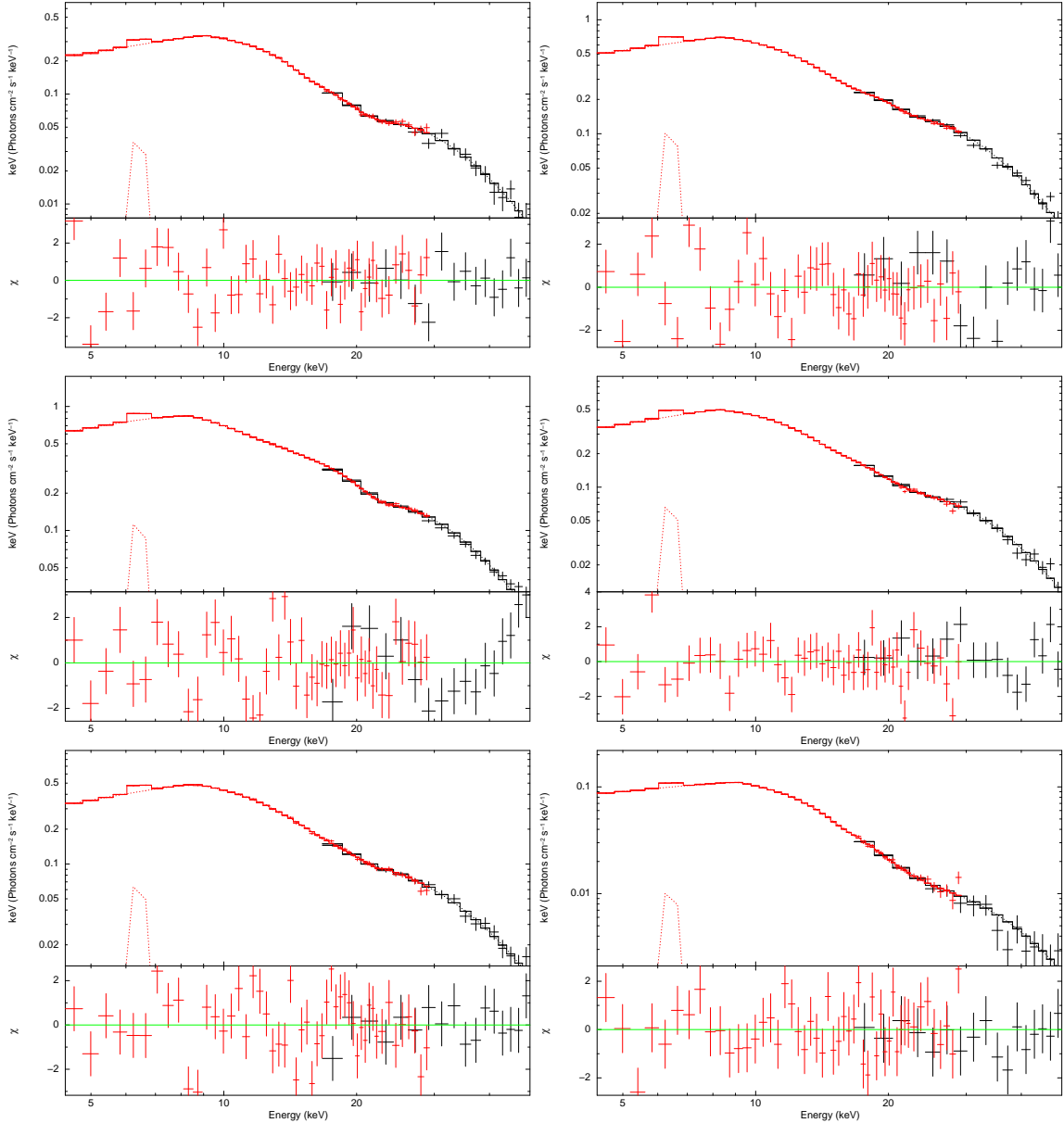

\centering
\includegraphics[angle=270,width=0.425\textwidth]{spectrum/xh93032010101_euf.ps}
\includegraphics[angle=270,width=0.425\textwidth]{spectrum/xh93032010206_euf.ps}
\includegraphics[angle=270,width=0.425\textwidth]{spectrum/xh93032010301_euf.ps}
\includegraphics[angle=270,width=0.425\textwidth]{spectrum/xh93032010406_euf.ps}
\includegraphics[angle=270,width=0.425\textwidth]{spectrum/xh93032010500_euf.ps}
\includegraphics[angle=270,width=0.425\textwidth]{spectrum/xh93032160201_euf.ps}
\caption{The hard X-ray spectra of 4U 0115+63 in the range of 4 -- 50 keV obtained by PCA (the black line) and
 HEXTE (the red line) during the observation in pointings 93032010101, 93032010206.
 93032010402, 93032010406, 93032010500, and 93032160201.}
\end{figure*}

For the spectra combined with JEM-X and IBIS, around the peak of the outburst, the lower band (3 -- 4 keV) of spectrum is a little lower than the continuum model spectrum if we only use the power law model with high energy cut off to fit the
continuum spectra. Hence, we add the column absorption component in the fittings. But for the PCA's data,
because we ignore the energy channels below 4 keV, where photo-electric absorption is not prominent
in the low energy band, therefore column density absorption is not included.

The spectra of 4U 0115+63 during the outburst obtained by JEM-X and IBIS are shown in Figure 5. The fitted spectral parameters are presented in Table 5. For Rev 675, the detection significance level of the source by JEM-X and IBIS is low ($<7\sigma$), no line features could be found in the spectrum, so we do not include it in the following analysis. For the spectrum in the range of 3 -- 100 keV by INTEGRAL, We observed cyclotron resonance absorption features up to the 5th harmonic resonance energy. The fundamental line energy is determined at 10 -- 15 keV in different epoches of the outburst. The second harmonic energy is around 22 keV, and the third harmonic energy is about 34 keV, and the fourth harmonic energy is around 42 keV. The fifth harmonic energy has a large uncertainty which varies from 49 keV to 55.5 keV, but the F-test shows this component might not be so significant (probabilities are much higher than 0.01), so we do not include the 5th absorption line in the spectral fittings. We also listed the probability of F-test,
The result obtained through PCA and HEXTE are presented in Table 6 \& Fig. 6, only the first and second harmonic lines are detected.
During our fitting, we found that the uncertainties of third cyclotron lines are very large.
So we fixed the width at what we got when this parameter is freely fitted, which is 1.0, in order
to increase the confidence of the fitting. The position of the third line is always around 35 keV, which is not influenced by the fixed
width.

In addition, in some spectra obtained by IBIS and JEM-X, the iron line at 6.4 keV is too weak to
convince us to add this feature into the spectra. Previous authors have found that the iron emission line at 6.4 keV is not neglectable in the spectral fittings (Tammura et al. 1992, Nakajima et al. 2006; Mihara et al. 2004 and Tsygankov et al. 2007). Here
we have adopt the iron line at 6.4 keV with fixed width at 0.2 keV just as Tsygankov et al. (2007)
have done. It doesn't matter whether we initially add this additive component into our model, because
if the iron line is not obvious, then it would be too weak to significantly affect the fitting result.

In Figure 6, the hard X-ray spectra of 4U 0115+63 from 4 -- 50 keV with RXTE/PCA and RXTE/HEXTE are presented. Here, we select 6 spectra from
the well fitted spectra of all observational pointings. Indeed all of the spectra obtained by RXTE have shown prominent
iron emission line at 6.4 keV. Since data of RXTE only covers the energy range below 50 keV and spectral data points are not so significantly above 30 keV, we only detected the cyclotron absorption line features at fundamental line energy around 10 --15 keV and the second harmonic energy at 22 keV.

\section{DISCUSSION}

\subsection{Spin-up Torque Versus Luminosity}

This source is observed continuously during the giant outburst in 2008, which enable us to study the spin properties and accretion physics in details. In \S. 3.1, we have fitted the data points to derive the spin period, and its derivative and orbital parameters. We found that the neutron star still undergone the spin-up process during the 2008 outburst. In the fittings, we have assume a constant spin-up torque $\dot P$. However, in the standard accretion model, the spin-up torque should correlate to its accretion luminosity.
 If we could build a simplest picture of the disk accretion, material is captured to magnetic field lines at $r_m$ and transported
to the magnetic poles. We assume all the angular momentum carried by the material is transferred to the neutron star, for a Keplerion disk, the neutron star would have a spin up toque as

\begin{equation} \label{eq:N}
N = \dot{M}\sqrt{GM_xr_m}.
\end{equation}

A neutron star with moment of inertia $I_x$ would subject to the torque in Equation (\ref{eq:N}),
when enduring outburst with high luminosity, its spin up rate is given by

\begin{equation} \label{eq:2pinu}
2\pi I_x\dot{\nu} = (GM_x)^{3/7} \mu^ {2/7}\dot{M}^{6/7}.
\end{equation}

Assuming the exhibited energy is from accretion, then all the gravitational potential energy of the accreted material would transport
to the neutron star's surface, with luminosity

\begin{equation} \label{eq:Lx}
L_x \approx G\dot{M} M_x /r_x
\end{equation}

Here $r_x$ is the radius of neutron star. Hence we could deduce

\begin{equation} \label{eq:nuL}
\dot{\nu} = \alpha L_x^{6/7},  \alpha = \frac{\mu^{2/7}r_x^{6/7}}{2\pi I_x(GM_x)^{3/7}}
\end{equation}


If we adopt the parameter of neutron star as follows, radius $r_x = 10\,km$, mass $M_x = 1.4 M_{\sun}$,
moment inertia $I_x = 2M_x r_x^2/5$. The magnetic field of the neutron star could be estimated through
cyclotron absorption energy we've observed. The difference between the first and second harmonic cyclotron
energy is 10 keV, if we adopt $\Delta E \approx 10\, keV B_{12}(1.2/1+z), z = G_xM_x/r_x c^2$, hence z$\sim$0.2
and B$\sim10^{12}~G$. Then we could get the value of $\alpha\sim 3.348\times 10^{-44}$.

Additionally, we could estimate the value of $\alpha$ using the observed data during the outburst in 2008. We can integrate Equation (\ref{eq:nuL}) into Equation (\ref{eq:Pspin}), here the
 $\dot{P} = -\frac{\dot{\nu}}{\nu^2} = -P^2 \dot{\nu} = \alpha P^2 L^{6/7}$, then we fit the function with the data when the luminosity
 is declining from the peak. The measured $\alpha = (1.918 \pm 0.339)\times 10^{-44}$ is still similar to though is a little lower than the theoretical value, suggesting that the accretion model described above could well predict the spin-up processes of this neutron star system.
It should also be pointed out that there still exist some uncertainties in estimating $\alpha$. In observations, the distance of the source and the surface magnetic field of the neutron star still have uncertainties. In theory, the Eq. (7) assumes that the whole accretion power has been transported to X-ray luminosity, but in realty only part of the accretion power could transport to X-ray luminosity. Anyway, using the observed data during the giant burst of 4U 0115+63, we have firstly confirmed the relation of spin-up torque versus luminosity in this source.

\subsection{Apsidal Motion}

In this paper, we derived the apsidal motion of the binary system by comparing different measurements.
The apsidal motion can probe the gravitational field and the internal structure of the companion star.
The methods and equations (9 $\sim$ 11) follow other authors, e.g. Kelley et al. (1981) who discussed
the apsidal motion in details, and Raichur \& Paul (2010) measured three sources' apsidal motion.
They assume the spin angular momentum of the companion star is parallel to the orbital angular momentum.
 $\dot{\omega}$ is related to the apsidal motion constant and the rotational velocity of the companion star as (Kelley et al. 1981)

\begin{equation} \label{eq:omega}
\dot{\omega} = \frac{2\pi k}{P_{orb}} (\frac{R_c}{a})^5 [15q f(e) + \Omega ^2 (1+q)g(e)],
\end{equation}

with
\begin{equation} \label{eq:omega}
f(e) = (1+\frac{3}{2} e^2 + \frac{1}{8} e^4)(1 - e^2)^{-5},
\end{equation}

\begin{equation} \label{eq:omega}
g(e) = (1 - e^2)^{-2}
\end{equation}

Here $q$ is the ratio of the mass of neutron star to the mass of its companion, $R_c$ is the radius of the companion star,
$\Omega$ is the ratio of rotational angular velocity of the companion to the orbital frequency $(2\pi / P_{orb})$, $a$ is the
full semimajor axis of the orbit. The apsidal motion constant $k$ describes the companion's internal matter density distribution.
We adopt the parameter suggested by Vacca et al (1996) and Negueruela \& Okazaki (2001), The companion, mass of V635 Cas is about $19M_{\sun}$, with radius
at $8R_{\sun}$, $i = 43^{\circ}$, rotational velocity $\nu \sin\,i = 290 \pm 50\,km\,s^{-1}$, hence we could deduce $k = 0.0025 \pm 0.0003$.
However, there are 70\% uncertainty in mass and 30\% uncertainty in radius of the companion star. If we include these uncertainties,
it would lead to a larger error bar of 0.0041. Anyway the apsidal motion of the binary can put the constraint on the mass-radius relation of the companion star. We derive $\dot\omega \sim 0.048^\circ \pm 0.003^\circ yr^{-1}$ which is consistent with the value by Tamura et al (1992) and
Raichur \& Paul (2010).

\subsection{The Resonance Energy and Other Spectral Parameters}

In \S 3, we have obtained all the fitted spectral parameters of the X-ray binary 4U 0115+63 from both INTEGRAL and RXTE's
data sets covering the outburst of 4U 0115+63 in 2008 April -- May, when this source's
X-ray luminosity changed from $(1.28 - 10.90) \times 10^{37} {\rm ergs\,s^{-1}}$ in the range of 5 -- 50 keV, assuming the distance is 7 kpc. In this section, the possible correlations between cyclotron absorption line parameters and continuum spectral parameters will be approached. Specially, the variations of the fundamental line energy during the giant outburst and the possible physical origins will be briefly discussed.

It is still hard to tell which factors could determine the cyclotron feature or the continuous spectra.
In order to get an overview of all the related parameters in spectral fittings, we first present some parameters' variations with observed time
during this giant outburst.

\begin{figure}
\centering
\includegraphics[angle=0,width=8cm]{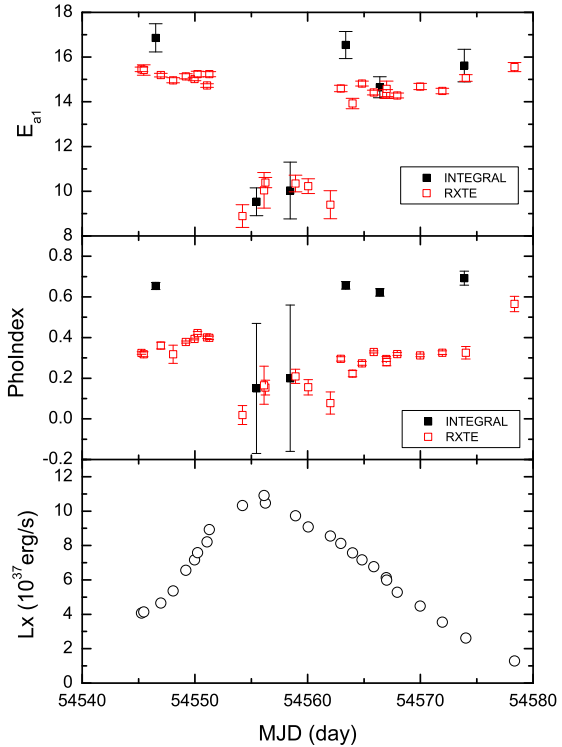}
\caption{The evolution of the fundamental resonance energy and photo index with time by both
INTEGRAL (blank squares) and RXTE (solid squares) observations. The sudden changes of both parameters are observed from the transition in the rising phase to the peak phase and in the peak to decreasing phase.}
\end{figure}

In Figure 7 we showed the variations of the fundamental line energy $E_{a1}$ and the photo index $\Gamma$ with observed time. The luminosity changes during the outburst are also displayed (bottom panel of Fig. 7), where we can clearly see the rising phase, the peak phase and deceasing phase of the whole outburst. In the plots, we have shown both the spectral results by INTEGRAL and RXTE. The mean photon index $\Gamma$ is different for the two instruments during the rising and decreasing phases: $\Gamma\sim 0.6$ for INTEGRAL data and $\Gamma\sim 0.4$ for RXTE. This effect may be due to the possible different observing off-angle of the two satellite when observing the outburst and there are more absorption lines detected in INTEGRAL's spectrum than RXTE, which would also make INTEGRAL's luminosity a little lower than RXTE's.

Anyway, the interesting correlation and transition are discovered in our observed results. $\Gamma$ and $E_{a1}$ vary nearly simultaneously, larger values (softer spectrum) of $\Gamma$ at $E_{a1}\sim 15$ keV while lower values (harder spectrum) of $\Gamma$ at $E_{a1}\sim 10$ keV. This correlation is valid for both INTEGRAL and RXTE data sets (also see Fig. 8). More importantly, the sudden changes of both parameters occurred during the peak phase of the outburst, indicating that there would be a transition of this
process in the rising phase to the peak phase and in the peak to decreasing phase. In the rising phase and decreasing phase, $E_{a1}\sim 15$ keV with a softer spectrum; and in the peak phase, $E_{a1}\sim 10$ keV with a harder spectrum. We also noticed that in the spectral fitting of INTEGRAL data from 3 -- 100 keV, only during the peak phase, the significant column density absorption is detected (see Table 5).

This sudden changes of spectral property and fundamental line energy suggest that some physics results in this transition. 
The possible correlation between X-ray luminosity and $E_{a1}$ have been pointed out before. We also display the possible relation between X-ray luminosity and $E_{a1}$ in Fig. 9. As we mentioned previously, the derived X-ray luminosity from INTEGRAL data is generally lower than that from RXTE data (by $\sim 20\%$). For each instrument's results, we can see the possible correlation: higher luminosity with lower values of $E_{a1}$. But the sudden transition of the $E_{a1}$ changes are clearly found rather than a real correlation function. For INTEGRAL's data sets, when $L_x > 6.76\times 10^{37} \rm ergs\,s^{-1}$, we detect the familiar fundamental CRSF at
$E_{a1} \approx$ 10 keV, below this luminosity level, the fundamental line energy changes to be $E_{a1} \approx$ 15 keV; while as for RXTE's data sets, the criterion is around $L_x > 8.5 \times 10^{37} \rm ergs\,s^{-1}$. For RXTE data points, we also noticed that for the similar X-ray luminosity ranges (from $(8.4-9)\times 10^{37}$ ergs s$^{-1}$), both two fundamental line energies at $\sim 10$ keV and $\sim 15$ keV are detected, suggesting that the X-ray luminosity may be not the unique factor in determining the fundamental line energy variations.
This transition is not complete yet because we do not get enough
data points when the fundamental line energy distributes from $E_{a1}\sim 12-14$ keV during the sudden transition phases.

\begin{figure}
\centering
\includegraphics[angle=0,width=8.2cm]{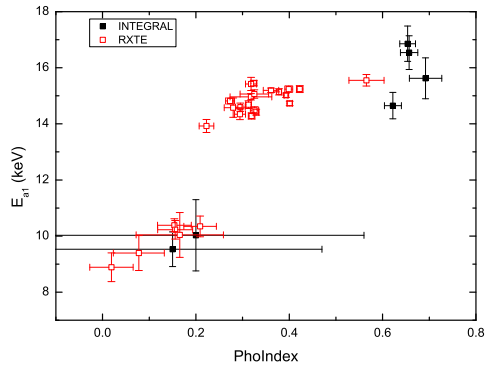}
\caption{The relationship between photo index and fundamental cyclotron energies
(data points obtained from both INTEGRAL and RXTE). }
\end{figure}

\begin{figure}
\centering
\includegraphics[angle=0,width=8.2cm]{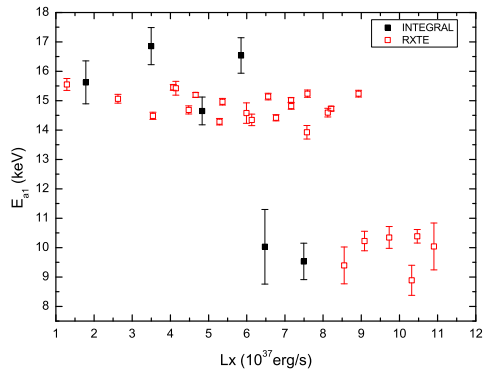}
\caption{The relationship between X-ray luminosity and fundamental cyclotron energies
(data points obtained from both INTEGRAL and RXTE). }
\end{figure}

Burnard et al. (1991) studied this luminosity dependent relationship, 
and assumed that the CRSF is formed above
the surface of neutron star at height $h_r$ in the accretion column. Based on the dipole law
of the magnetic field, the fundamental cyclotron absorption should follow the equation given below:
\begin{equation} \label{eq:Ea1}
E_{a1} \propto (R_{NS} + h_r)^{-3}(1 + z_g)^{-1},
\end{equation}
where $R_{NS}$ stands for the radius of the neutron star, $z_g$ denotes for the gravitational redshift.
Here we could treat the redshift item as a coefficient, for simplicity, we set $z_g$ = 0. Then we could
describe the influence of height on resonance energy in a convenient way.
\begin{equation} \label{eq:ratio}
\frac{h_r}{R_{NS}} \approx (\frac{E_a}{E_0})^{-1/3} - 1.
\end{equation}

Burnard et al. (1991) have given their estimation:
\begin{equation} \label{eq:hr}
h_r \approx \frac{L_x}{L^{eff}_{Edd}H_\bot}R_{NS}.
\end{equation}
$L^{eff}_{Edd}$ is the Eddington luminosity along the magnetic field, identical to the conventional Eddington
luminosity for a neutron star with a mass of 1.4$M_\odot$, $L^{eff}_{Edd} = 2.0\,\times10^{38} ergs\,s^{-1}$. $H_\bot$ is
the ratio of the Thomson cross section to the Rosseland-averaged electron scattering cross section for radiation flows across the magnetic fields. Here we adopt $H_\bot$ = 1.23 deduced by Mihara et al. (2004).

So we could employ the $E_{a1}$ into the above equations, then estimate $h_r/R_{NS}$ as a function
of X-ray luminosity from our observations by INTEGRAL and RXTE. The results are shown in Figure 10. Here we assume $E_0$ = 18 keV, which is the fundamental resonance energy at the surface of the neutron star, taken to be the possible maximum value of our observations. We
 deduce all the $h_r/R_{NS}$ according to equation (13). However we did not find the similar results
done by Mihara et al. (2004). Even we have treated data sets from INTEGRAL and RXTE respectively, our results could not
confirm this theoretical model.

\begin{figure}
\centering
\includegraphics[angle=0,width=0.425\textwidth]{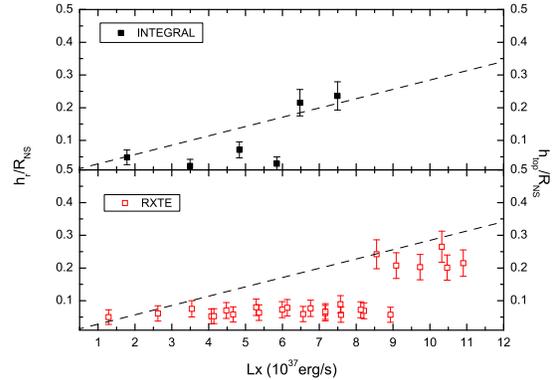}
\caption{The height, converted from equation (13) versus luminosity. The dashed line represents
the column top height, calculated by equation (14).}
\end{figure}

Instead of this linear relation ship suggested as equation (14), we find a phase transition similar to $\Gamma$ and $E_{a1}$. The trigger of the sudden transition may closely related to the luminosity of the source. As the luminosity increases above a certain level,
condition might change suddenly instead of what equation (14) has suggested. Besides, as shown in Figure 10, the  $h_r/R_{NS}$ ratio drops suddenly when it decreases to under 0.1 level, this trend could also be seen in the work done by Nakajima et al. (2006). So the mechanism of the fundamental cyclotron
absorption may be affected by other unknown physical factors, or the magnetic field at the surface of neutron star might not be consistent with the dipole assumption.

We also study the ratios $E_{a1}/E_{a2}$ and $h_r/R_{NS}$ versus time of the outburst, as shown in Figure 11. Since $E_{a1}$ changes significantly during the outburst, while $E_{a2}$ is relatively stable around 22 keV in the whole outburst, the variation of the ratio $E_{a1}/E_{a2}$ shows the similar behavior to the evolution of $E_{a1}$: in the rising phase, the ratio is about 1.3, in peak phase the ratio switches to be around 2, and in decreasing phase, the ratio returns to be near 1.3. The transition may be related to the X-ray luminosity level, deserved more detailed studies.

\begin{figure}
\centering
\includegraphics[angle=0,width=0.425\textwidth]{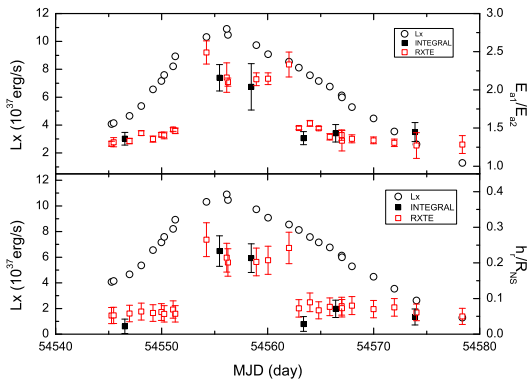}
\caption{The evolution of $Ea_{2}/Ea_{1}$ and $h_r/R_{NS}$ with time of the outburst. The circles represent the luminosity of the outburst. We could see the sudden changes at the transition from rising phase to peak phase in MJD 54554, and transition from peak phase to decreasing phase in MJD 54562. }
\end{figure}

Coburn et al. (2002) found that there exists a positive correlation
between absorption depth and $W/E_a$ in phase averaged spectra of the X-ray pulsar. This indicates
that this source has a tall cylindrical shape accretion column rather than a flat
coin shape, suggested by Kreykenbohm et al. (2004). In Figure 12 we presented the
ratio of the width to energy of both $E_{a1}$ and $E_{a2}$ versus
the depth. The positive correlation between $W/E_a$  and depth is found for both two lines, though large scattering and uncertainties still exist. This result, along with what Kreykenbohm et al. (2004) and Nakajima et al. (2006) have obtained, strengthens the cylindrical column geometry assumption.

\begin{figure}
\centering
\includegraphics[angle=0,width=0.45\textwidth]{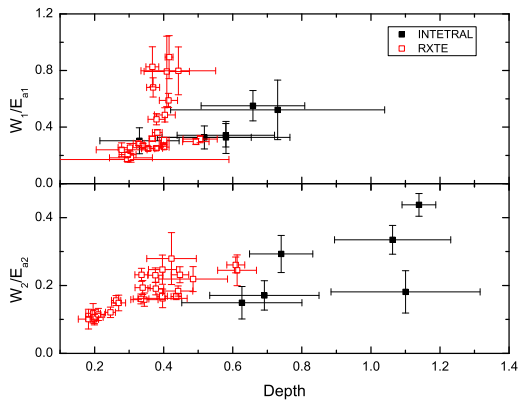}
\caption{$W/Ea$ of fundamental and second harmonic resonance
energy of 4U0115+63, varies with their energy values, respectively. }
\end{figure}

We have fitted all the continuum spectra with a power-law model plus a high energy cutoff. The photon index variations have been plotted before. Finally, we presented the variations of other two spectral parameters $E_{cut}$ and $E_{fold}$ with time during the outbursts in Fig. 13. Most values of $E_{cut}$ do not change too much during the whole outburst, it stays around 8.5 keV for both RXTE and INTEGRAL observations, which is consistent with the results by Tsygankov et al (2007) and
Makishima et al (1999). For $E_{fold}$, INTEGRAL results show a higher values
($\sim 12$ keV) than that ($\sim 8$ keV) obtained by RXTE, which would be due to the wider energy range by INTEGRAL which covers higher ranges up to 100 keV.

\begin{figure}
\centering
\includegraphics[angle=0,width=8cm]{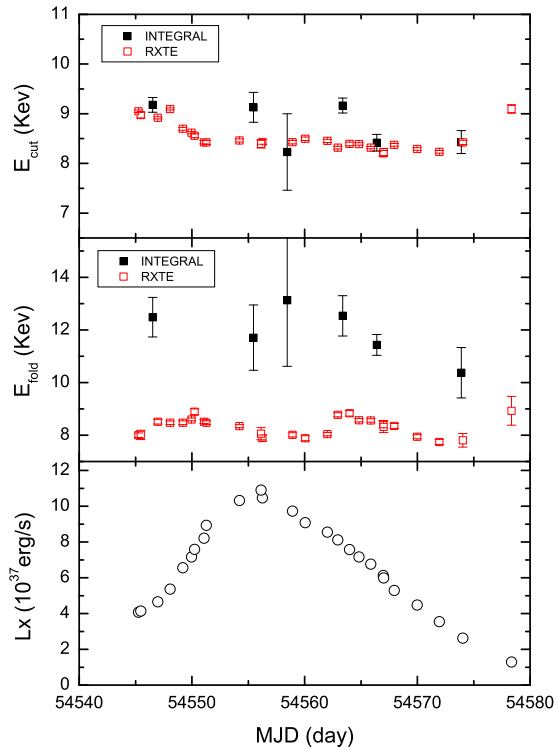}
\caption{The evolution of cut-off energy and fold energy with time during the outburst (data points from both INTEGRAL and RXTE).}
\end{figure}

\section{Summary and Conclusion}

The giant outburst in neutron star high mass X-ray binaries provides us a good chance to study the accretion physics and interaction between magnetosphere and accretion materials. We observed the giant outburst of 4U 0115+63 during April -- May 2008 using both INTEGRAL and RXTE data, performing both timing and spectral analysis in details. The spin period is determined to be
$3.61430\,\pm\,0.00003$ s at MJD 54566, and we confirm that this source is enduring spin up state at rate of
$\dot{P}_{\rm spin} = (-7.24 \pm 0.03)\times10^{-6}$ s d$^{-1}$ during the outburst. In addition, we determined the angle of periapsis of the orbit $\omega = 48.67^\circ \pm 0.04^\circ$ during the 2008 outburst,
as we compare with previous works done by other authors, we get $\dot{\omega} = 0.048^\circ \pm 0.003^\circ$ yr$^{-1}$. We also noticed the pulse profile's merging behavior during the outburst.

Furthermore, spectral properties of 4U 0115+63 during the different phases of the outburst are studied. All spectra from both INTEGRAL and RXTE
 are fitted with the power-law model plus a high energy cutoff and added by two components, cyclotron resonant absorption features and a gaussian
line at 6.4 keV. We have searched for the cyclotron resonant absorption features up to 5th harmonic cyclotron line with INTEGRAL data.
The fundamental line energy varies during the outburst at 10 keV or 15 keV respectively; the second harmonic energy at 22 keV;
the third at $\sim 34$ keV; 4th at $\sim 45$ keV. The fifth harmonic energy is not significantly detected.
Hence we have confirmed that 4U 0115+63 shows the four significant cyclotron absorption lines features.

The present observations with INTEGRAL and RTXE provide us an opportunity to study variations of all spectral parameters
in this source with time and resonance energy during the 2008 giant outburst.
A positive relationship between photo index and fundamental cyclotron energy is found (see Figure 8). The evolutions of fundamental cyclotron energy and spectral properties with time during the outburst show the sudden transitions: at the rising phase, the source shows the soft spectrum with $E_{a1}\sim 15$ keV; in the peak phase, $E_{a1}\sim 10$ keV with a harder spectrum, and high column density; then into the decreasing phase, returning to the soft spectrum with $E_{a1}\sim 15$ keV. The
 luminosity dependence of resonance fundamental energy is also indicated in our observations, but in the similar luminosity range of $\sim (8.4-9)\times 10^{37}$ ergs s$^{-1}$, both two fundamental energies at $\sim 10$ and 15 keV are detected. So the physical parameters other than the single X-ray luminosity may play a role in determining the spectral properties during the outburst. Instead of a positive relationship of $h_r/R_{NS}$ and luminosity as expected, we also found a transition which makes a sudden change for the ratio (below $\sim 0.1$ or around 0.2), similar to fundamental line energy, which indicates that the environment near the surface of neutron star is much more complicated than we originally thought.
This transition process during the outburst may require further studies both in observational and theoretical work. The X-ray luminosity should be a important parameter, when it reaches a certain critical value, sudden changes of accretion and radiation could occur or other physical conditions would change. Another positive correlation between depth and $W/E_a$ ratio is consistent with previous work, which supported the tall cylindrical
geometry assumption of the accretion column. The distinct spectral properties (different photon indices and fundamental line energies) between the peak phase and the rising/decreasing phases suggested different accretion physics and environments in the peak luminosity ranges.

\section*{Acknowledgments}

We are grateful to the referee for the fruitful comments and suggestions to improve the manuscript. This paper is based on observations of INTEGRAL, an ESA's project with instrument
 and science data center funded by ESA member states (principle investigator
 countries: Denmark, France, Germany, Italy, Switzerland and Spain),
 the Czech Republic and Poland, and with participation of Russia and US.
Data of RXTE obtained from the High Energy Astrophysics Science Archive Research
 Center (HEASARC), provided by NASA's Goddard Space Flight Center. The work is supported by the
National Natural Science Foundation of China under grants
10803009, 10833003, 11073030.

\bsp

\label{lastpage}

\end{document}